\documentclass{jmlr}

\jmlrproceedings{}{}

\usepackage{longtable}

\makeatletter
\let\Ginclude@graphics\@org@Ginclude@graphics 
\makeatother

\title[Policy Iteration for Two-Player General-Sum Stochastic Stackelberg Games]{Policy Iteration for Two-Player General-Sum Stochastic Stackelberg Games}

\usepackage[T1]{fontenc}
\usepackage{graphicx}
\usepackage{booktabs}
\usepackage[misc]{ifsym}

\usepackage{cite}

\usepackage{tikz}
\usetikzlibrary{arrows,automata}

\usepackage{mathtools}

\usepackage[textsize=tiny]{todonotes}

\usepackage{algorithm}
\usepackage{algorithmic}

\usepackage[capitalize,noabbrev]{cleveref}

\usepackage{comment}
\usepackage{url}

\providecommand{\argmax}{\operatornamewithlimits{argmax}} 
\providecommand{\R}{\mathbb{R}} 
\providecommand{\N}{\mathbb{N}} 
\providecommand{\E}{\mathbb{E}} 
\providecommand{\x}{\times}
\providecommand{\where}{\mathrm{where~}}

\providecommand{\eqs}{\stackrel{s}{=}}
\renewcommand{\geq}{\geqslant} 
\renewcommand{\leq}{\leqslant} 

\DeclarePairedDelimiterX{\inner}[2]{\langle}{\rangle}{#1, #2}
\DeclarePairedDelimiter{\norm}{\lVert}{\rVert}
\DeclarePairedDelimiter{\abs}{\lvert}{\rvert}

\newcommand{\alglinelabel}{%
  \addtocounter{ALC@line}{-1}
  \refstepcounter{ALC@line}
  \label
}

\begin{document}



\author{\Name{Mikoto Kudo}\Email{mikoto@bbo.cs.tsukuba.ac.jp}\\
\Name{Youhei Akimoto}\Email{akimoto@cs.tsukuba.ac.jp}\\
\addr University of Tsukuba \& RIKEN AIP, Tsukuba, Ibaraki 305-8577, Japan
}



\maketitle

\begin{abstract}
We address two-player general-sum stochastic Stackelberg games (SSGs), where the leader’s policy is optimized considering the best-response follower whose policy is optimal for its reward under the leader.
Existing policy gradient and value iteration approaches for SSGs do not guarantee monotone improvement in the leader’s policy under the best-response follower.
Consequently, their performance is not guaranteed when their limits are not stationary Stackelberg equilibria (SSEs), which do not necessarily exist.
In this paper, we derive a policy improvement theorem for SSGs under the best-response follower and propose a novel policy iteration algorithm that guarantees monotone improvement in the leader’s performance. 
Additionally, we introduce Pareto-optimality as an extended optimality of the SSE and prove that our method converges to the Pareto front when the leader is myopic.
\end{abstract}


\section{Introduction}
A Markov decision process (MDP) is a mathematical framework that models the decision-making of agents in dynamic environments, which serves as the foundation of reinforcement learning (RL). 
The performance of the agent's policy is evaluated by assessing the state-action sequences it produces with the reward function.
This performance is quantified by the expected cumulative discounted reward.
An optimal policy is one that maximizes this expected return.

One extension of MDPs for multi-agent systems is a stochastic game, also known as a Markov game, where multiple agents with individual reward functions simultaneously attempt to develop policies that maximize their expected cumulative discounted rewards within a single environment.
Stochastic games were initially proposed in game theory, and the solution is defined as a tuple of policies that form a specific type of equilibrium, such as a Nash equilibrium. 
The problem setting is also characterized by the relationship between the reward functions, such as the zero-sum (i.e., competitive) and fully cooperative settings.

In this study, we addressed the problem of determining stationary Stackelberg equilibria (SSEs) in 2-player stochastic games with general relationships in the rewards.
An SSE is defined as a tuple of stationary policies of two types of agents, \emph{leader} and \emph{follower}, such that the leader's policy maximizes the leader's payoff when the follower takes the policy that is the best response to the leader's policy with respect to the follower's payoff.
We assume that the follower always takes the best response to any leader's policy. We refer to such a follower as a \emph{best-response follower}.
Additionally, we assume that the follower's best responses are always computable to the leader and that the environment is known. 
No other assumptions are made on the follower's algorithm.

This situation arises, for example, in the design of e-commerce platforms aimed at the maximization of the site owner's profit.
The site user (follower) repeatedly makes purchasing decisions based on their preferences. In contrast, the site owner (leader) can configure various elements of the platform, such as page transitions, advertisements, and an incentive bonus.
Leveraging knowledge of the user's past behavior, the owner seeks to maximize long-term profit by anticipating the user's responses.
Assuming the user's responses are always optimal for their preferences and the owner's prediction is accurate enough, the resulting owner's problem can then be formulated as an SSG with a known best-response follower.

As an SSE does not always exist in the general-sum setting~\citep{Bucarey2022-ro}, the algorithm should converge a leader's policy to the SSE if it exists; otherwise, it must converge to a policy that achieves reasonably satisfactory performance.
However, existing methods based on dynamic programming operators \citep{Bucarey2022-ro,Zhang2020-hw} and policy gradient methods with total derivatives \citep{Zheng2022-xj,vu2022stackelberg} do not guarantee the convergence to the SSEs in our setting, even if they exist. 
Moreover, these algorithms may converge leaders’ policies to low-quality ones because they do not guarantee monotone improvement of the leaders’ performance under the best-response follower.
In this work, we focus on policy improvement methods and develop an algorithm that satisfies the above requirements.

\subsection{Main Contributions}
Our contributions are summarized as follows.
First, we explicitly derive the fixed point of the dynamic programming operator used in \citet{Bucarey2022-ro,Zhang2020-hw}.
This result reveals that the fixed point does not necessarily bring a reasonable leader policy when it is not guaranteed to be an SSE (\cref{sec:BucareyMethod}).
Second, a \emph{policy improvement theorem} for general-sum stochastic Stackelberg games with the best-response follower is derived (\cref{sec:improvement}), based on which a novel policy iteration algorithm is proposed (\cref{sec:approach}). 
Finally, we introduce the concept of \emph{Pareto-optimality} as an extended optimality of the SSE.
Policies that realize the Pareto-optimal value functions or their neighborhood with arbitrary precision always exist. 
Moreover, Pareto-optimality agrees with the SSE it exists (\cref{sec:po_policy}).
Then, we prove that the proposed method monotonically improves the state values toward the Pareto front and converges to the front when the leader is myopic (\cref{sec:approach}).
To the best of our knowledge, this is the first theoretical guarantee in general-sum myopic-leader SSGs.

\section{Preliminaries}\label{sec:preliminaries}

\subsection{Markov Decision Process}

An MDP is a stochastic process with rewards in which the state transitions depend on the actions of an agent.
An MDP is defined as a tuple $(\mathcal{S},\mathcal{A},p,\rho,r,\gamma)$: a finite state space $\mathcal{S}$, a finite action space $\mathcal{A}$, a transition function $p: \mathcal{S}\x\mathcal{A}\to \Delta(\mathcal{S})$, an initial state distribution $\rho\in\Delta(\mathcal{S})$, a bounded reward function $r: \mathcal{S}\x\mathcal{A}\to\R$, a discount rate $\gamma\in[0,1)$, where $\Delta(\mathcal{S})$ is a set of probability distributions on $\mathcal{S}$.

The agent determines its action following its Markov policy. 
Let $\mathcal{W}:=\{f:\mathcal{S}\to\Delta(\mathcal{A})\}$ be a set of (stochastic) decision rules. Then, at the time step $t\in\N$, the agent has the decision rule $f_t\in\mathcal{W}$ and selects the action $a_t\in\mathcal{A}$ in the current state $s_t\in\mathcal{S}$ as $a_t\sim f_t(s_t)$. 
The conditional probability is denoted as $f_t(a_t|s_t)$.
If the decision rule is invariant over time ($f_t=f$), the Markov policy is called a \textit{stationary policy} and is simply denoted by $f\in\mathcal{W}$.

The next state $s_{t+1}$ is stochastically determined as $s_{t+1}\sim p(s_{t+1}|s_t,a_t)$ when the agent selects its action $a_t$ in the state $s_t$.
Repeating such transitions with one stationary policy $f\in\mathcal{W}$ generates the state-action sequence $\{(s_t,a_t)\}_{t=0}^{\infty}$ under $f$.
Then, the performance of $f$ can be defined by evaluating the state-action sequence with the reward function, and the optimal policies are defined as policies with the highest performance.
Optimal policies are defined as stationary policies that maximize the expectation of the cumulative discounted reward under $f\in\mathcal{W}_A$ conditioned by an initial state $s\in\mathcal{S}$, which is given by
\begin{equation*}
	V^{f}(s):=\E^{f}\left[\sum_{t=0}^{\infty} \gamma^tr(s_t,a_t)\middle| s_0=s\right],
\end{equation*}
for all $s\in\mathcal{S}$, 
where $\E^{f}\left[\cdot|s_0=s\right]$ is the expectation over $\{(s_t,a_t)\}_{t=0}^{\infty}$ generated under $(f, p)$ given initial state $s$.
$V^{f}(s)$ is called a state value function of stationary policy $f$.
Hereafter, we consider only stationary policies and refer to them simply as policies.

There always exists a deterministic optimal policy~\citep{Sutton2018-jf} such that
\begin{equation}
	f^*(s) \in\argmax_{a\in\mathcal{A}} \bigg\{r(s,a) +\gamma\E_{s'\sim p(s'|s,a)}\left[\max_{f\in\mathcal{W}}V^{f}(s')\right]\bigg\}\label{eq:optPolicy_MDP}
\end{equation}%
for all $s\in\mathcal{S}$.
The main algorithms used to determine the optimal policy are value iteration and policy iteration when the transition and reward functions are known, and RL when they are unknown~\citep{Sutton2018-jf}.

\subsection{Stochastic Game and Stackelberg Game}

Stochastic games~\citep{Shapley1953-me,Solan2022-ce}, also known as Markov games, are an extension of MDPs for multi-agent settings. 
When there are two agents, A and B, they have finite action spaces $\mathcal{A}$ and $\mathcal{B}$, bounded reward functions $r_A: \mathcal{S}\x\mathcal{A}\x\mathcal{B}\to\R$ and $r_B: \mathcal{S}\x\mathcal{A}\x\mathcal{B}\to\R$, discount rates $\gamma_A\in[0,1)$ and $\gamma_B\in[0,1)$ and sets of policies $\mathcal{W}_A:=\{f:\mathcal{S}\to\Delta(\mathcal{A})\}$ and $\mathcal{W}_B:=\{g:\mathcal{S}\to\Delta(\mathcal{B})\}$, respectively.
The state $s\in\mathcal{S}$ is shared among the agents, and the transition function is defined as $p: \mathcal{S}\x\mathcal{A}\x\mathcal{B}\to\Delta(\mathcal{S})$; that is, the transition probability depends on the actions of all the agents.

The expectation of the cumulative discounted reward under stationary policies $f\in\mathcal{W}_A$ and $g\in\mathcal{W}_B$ is defined for $r_A$ and $r_B$, respectively.
Then, equilibrium policies, such as Nash equilibrium, can be identified as solutions to the problem by considering each cumulative reward as the agents' payoff.

The relationship of the reward functions between agents characterizes the game. 
In game theory, a game where $r_A=-r_B$ is called a zero-sum game, and the others are called non-zero-sum games. Both are called general-sum games.
In the context of a multi-agent RL, situations where $r_A \propto -r_B$ and $r_A \propto r_B$ are called competitive and cooperative, respectively.

The Stackelberg game is a game model in which a leader selects its strategy first, and then a follower selects its strategy that maximizes the follower's payoff against the leader's strategy.
This follower's strategy is called the \emph{best response} strategy against the leader's strategy. 
When the leader's strategy maximizes the leader's payoff under the follower's best response, the pair of strategies is called a Stackelberg equilibrium.
In this study, we consider the extension of Stackelberg games to stochastic games, such as the stochastic Stackelberg game (SSG).
Unlike the single-agent MDP, the optimal policy, \emph{SSE policy} defined below, is not guaranteed to exist, and even if it does, it is not necessarily deterministic.
An example scenario where the SSE policy does not exist is provided in the next section.

\section{Problem Setting}\label{sec:problem_setting}


A general-sum SSG is represented by $(\mathcal{S}, (\mathcal{A},\mathcal{B}), \hat{p}, \rho, (\hat{r}_A,\hat{r}_B), (\gamma_A,\gamma_B))$, where the subscripts $A$ and $B$ indicate the \emph{leader} and \emph{follower}, respectively.
The sets of policies are defined by 
\begin{align*}
	\mathcal{W}_A:=\{f:\mathcal{S}\to\Delta(\mathcal{A})\},
        \quad 
	\mathcal{W}_B:=\{g:\mathcal{S}\to\Delta(\mathcal{B})\},
\end{align*}
and the sets of deterministic policies are defined by
\begin{align*}
	\mathcal{W}_A^d:=\{f:\mathcal{S}\to\mathcal{A}\},
	\quad
	\mathcal{W}_B^d:=\{g:\mathcal{S}\to\mathcal{B}\}.
\end{align*}
Let $\mathcal{F}_{\mathcal{S}}:=\{v:\mathcal{S}\to\R\}$ be the set of real-valued functions with a state as input.   

For conciseness, we define the marginalization of $\hat{r}$ and $\hat{p}$ over the leader's action $a\in\mathcal{A}$ under an action distribution $f_s\in\Delta(\mathcal{A})$ as
\begin{align*}
	r_i(s,f_s,b)&:=\sum_{a\in\mathcal{A}}f_s(a)\hat{r}_i(s,a,b),\quad(\forall i\in\{A,B\})\\
	p(s'|s,f_s,b)&:=\sum_{a\in\mathcal{A}}f_s(a)\hat{p}(s'|s,a,b),
\end{align*}
where $s\in\mathcal{S}$ is a state and $b\in\mathcal{B}$ is a follower's action.
Hereafter, the players are indexed by $i \in \{A, B\}$. 

The state value functions of the leader and the follower are defined as follows.
\begin{definition}[State value function]
	Let $s\in\mathcal{S}$ be a state and $f\in\mathcal{W}_A,g\in\mathcal{W}_B$ be stationary policies of the leader and the follower.
	The state value functions $V_i^{fg}\in\mathcal{F}_{\mathcal{S}}$ of player $i\in\{A,B\}$ for a pair $(f, g)$ are given by
	\begin{equation*}
		V_i^{fg}(s):=\E^{fg}\left[\sum_{t=0}^{\infty}\gamma_i^t\hat{r}_i(s_t,a_t,b_t)\middle|s_0=s\right],
	\end{equation*}
	where $\E^{fg}\left[\cdot | s_0=s\right]$ is the expectation over the state-action sequence $\{(s_t,a_t,b_t)\}_{t=0}^{\infty}$ generated under $(f, g, \hat{p})$ given initial state $s$.
\end{definition}

The follower's best response is defined by employing the state value functions.
\begin{definition}[Follower's best response]
	Let $s\in\mathcal{S}$ be a state and $f\in\mathcal{W}_A$ a leader's policy.
	The set of the follower's best responses $R_B^*(f)\subseteq \mathcal{W}_B^d$ against the leader's policy $f$ is defined as 
	\begin{align*}
		R_B^*(f):=\{g\in \mathcal{W}_B^d~|~g(s)\in R_B^*(s,f)~\forall s\in\mathcal{S}\},
	\end{align*}
	where the best response action at state $s$ is defined as
	\begin{equation*}
		R_B^*(s,f):=\argmax_{b\in\mathcal{B}} r_B(s,f(s),b) + \gamma_B\E_{s'\sim p(s'|s,f(s),b)}\left[\max_{g\in \mathcal{W}_B}V_B^{fg}(s')\right].
	\end{equation*}
\end{definition}

As the optimal policies in single-agent MDPs are represented as in \cref{eq:optPolicy_MDP}, we can observe that the set of best responses $R_B^*(f)$ with fixed leader's stationary policy $f$ is the set of the optimal policies in terms of single-agent MDPs.
It is also shown that $R_B^*(f)$ is non-empty for all $f\in\mathcal{W}_A$.

We assume that the follower is a \emph{best-response follower}, meaning that it always adopts the best response to the leader's policy. 
If the leader adopts a stationary policy $f\in\mathcal{W}_A$, the best-response follower adopts the stationary policy $g\in R_B^*(f)\subseteq\mathcal{W}_B^d$.
This setup is essentially equivalent to a situation where we can only control the leader's policy, and the follower independently learns its policy with the ability to find the best responses in a finite amount of time for any leader's policy.
No other assumptions are made about the follower's learning process.

We assume that the set of the follower's best responses, $R_B^*(s,f)$, is a singleton for each $s\in\mathcal{S}$ and for each $f\in\mathcal{W}_A$. Otherwise, we break ties deterministically.
It follows that $R_B^*(f)$ is also a singleton for all $f\in\mathcal{W}_A$; thus, the follower's best response against $f$ is unique.
For simplicity, let $R_B^*(f)$ be the best response.
This enables us to define a state value function $V_A^{fR_B^*(f)}$ of $f$ under the best-response follower.
For simplicity, we denote
\begin{align*}
	V_A^{f\dag}:=V_A^{fR_B^*(f)}.
\end{align*}

We aim to determine an SSE policy, 
which is defined as the leader's stationary policy that maximizes the leader's state value function for all states under the best-response follower.
This policy and the SE value function are defined below.

\begin{definition}[SE value function]
	Let $s\in\mathcal{S}$ be a state.
	An SE value function is defined as
	\begin{align*}
		V_A^*(s) := \sup_{f\in\mathcal{W}_A} V_A^{f\dag}(s).
	\end{align*}
\end{definition}

\begin{definition}[SSE policy]\label{thrm:def_SSEP}
	If a leader's stationary policy $f^*\in\mathcal{W}_A$ satisfies $V_A^{f^*\dag}(s)= V_A^*(s)$ for all $s \in \mathcal{S}$, then $f^*$ is an SSE policy.
\end{definition}

\paragraph{Game without SSEs}
We provide an example game where SSEs do not exist. This example is first introduced by Jean-Marie in his seminar talk \citep{Jean-Marie2022-td} related to \citet{Bucarey2022-ro}.
We consider a game with two states $\mathcal{S} = \{s_1, s_2\}$, two leader's actions $\mathcal{A} = \{a_1, a_2\}$, and two follower's actions $\mathcal{B} = \{b_1, b_2\}$. 
The dynamics and the reward signals are deterministic and represented in \cref{fig:toytask}.
Let the leader's policy be denoted as $f_{s_1}(a_1) = p \in [0, 1]$ and $f_{s_2}(a_1) = q \in [0, 1]$. 

Given the leader's policy, it reduces to a standard MDP for the follower. 
The optimal value function for the follower is the solution to 
\begin{align*}
V_B^*(s_1) &= \max\{\gamma_B V_B^*(s_1), x + \gamma_B V_B^*(s_2)\} \cdot p + (-y + \gamma_B V_B^*(s_2)) \cdot (1 - p) ;
\\
V_B^*(s_2) &= \max\{\gamma_B V_B^*(s_2), x + \gamma_B V_B^*(s_1)\} \cdot q + (-y + \gamma_B V_B^*(s_1)) \cdot (1 - q) .
\end{align*}
There are three cases depending on the leader's policy $f$, summarized in \cref{tbl:sevalues}.
If $\gamma_B y > x$, we know that $(p, q) = (1, 0)$ leads to the case of $\gamma_B V_B^*(s_1) > x + \gamma_B V_B^*(s_2)$. Then, the maximum value of the leader's value at state $s_1$ is obtained as $V_A^{f\dag}(s_1) = \frac{1}{1 - \gamma_A}$. Similarly, we can check that $(p, q) = (0, 1)$ leads to the case of $\gamma_B V_B^*(s_2) > x + \gamma_B V_B^*(s_1)$. Then, the maximum value of the leader's value at state $s_2$ is obtained as $V_A^{f\dag}(s_2) = \frac{1}{1 - \gamma_A}$. 
We can see that there is a tradeoff between the values at $s_1$ and $s_2$ and two Pareto optimal points exist: $(V_A^{f\dag}(s_1), V_A^{f\dag}(s_2)) = (\frac{1}{1 - \gamma_A}, \frac{\gamma_A}{1 - \gamma_A})$ and $(\frac{\gamma_A}{1 - \gamma_A}, \frac{1}{1 - \gamma_A})$.
That is, there does not exist a single optimal policy that maximizes the values at all states simultaneously, implying that an SSE does not exist.


\begin{table}[t]
\centering
\caption{The follower's best response and the leader's value under the best response follower.}\label{tbl:sevalues}
\begin{tabular}{ccc}
    \toprule
    Case & $R_B^*(s_1,f), R_B^*(s_2,f)$ & $ V_A^{f\dag}(s_1)$,  $V_A^{f\dag}(s_2)$\\
    \midrule
    $\gamma_B V_B^*(s_1) > x + \gamma_B V_B^*(s_2)$ & $b_1, b_2$ & $\frac{p}{1 - \gamma_A}, \frac{p \gamma_A}{1 - \gamma_A}$ \\
    $\gamma_B V_B^*(s_2) > x + \gamma_B V_B^*(s_1)$ & $b_2, b_1$ & $\frac{q \gamma_A}{1 - \gamma_A}, \frac{q}{1 - \gamma_A}$ \\
    $\abs{ V_B^*(s_1) - V_B^*(s_2)} < x / \gamma_B$ & $b_2, b_2$ & $0, 0$ \\
     \bottomrule
\end{tabular}
\end{table}
\begin{figure}[t]
  \centering
  \begin{tikzpicture}[->,auto,node distance=12.5em, scale=1, transform shape]
    \tikzstyle{every state}=[fill=black!15,draw=black,text=black]
    \node[state] (A) {$s_1$};
    \node[state, right of=A] (B) {$s_2$};

    \path 
    (A) edge [loop left]  node {($a_1$, $b_1$) / (1, 0)} (A)
    (A) edge [left, above]  node {} (B)
    (B) edge [loop right] node {($a_1$, $b_1$) / (1, 0)} (B)
    (B) edge [left, below]  node[align=left] {($a_1$, $b_2$) / (0, $x$) \\ ($a_2$, *) / (0, $-y$)} (A);
  \end{tikzpicture}
  \caption{An example of a game without SSEs. 
  Transitions are deterministic and represented by arrows. 
  Each label shows (leader's action, follower's action)\,/\, (leader's reward, follower's reward), where $x > 0$ and $y > 0$.}
  \label{fig:toytask}
\end{figure}


\section{Related Work}\label{sec:related_works}


\citet{Bucarey2022-ro} proposed an algorithm for computing \emph{strong} SSEs in the 2-player general-sum stochastic games, where strong SSEs are introduced to break ties for the follower’s best response, which corresponds to a specific tie-breaking mechanism for the follower’s best response in our setting.
They provide sufficient conditions for their algorithm to converge to the SSEs.
However, there is little assurance of the leader's performance of the obtained policy when it does not converge to the SSEs or when there is no SSE. 
This problem is because the fixed point of the dynamic programming operator used in Bucarey et al. (2022) is different from the SE value function.
We discuss this in detail in \cref{sec:BucareyMethod}. 

A primary case where the algorithm of \citet{Bucarey2022-ro} converges and the fixed point coincides with the SE values is when the follower is myopic, i.e., when the follower maximizes its immediate rewards. 
This setting is adopted in some existing SSG studies.
For example, \citet{Zhao2023-jc} derived the upper bound of the regret of the leader's payoff in the cooperative tasks, where the reward function is shared between the leader and the follower, with information asymmetry under the myopic follower.
\citet{Zhong2023-fn} proposed an algorithm in the SSG where the group of myopic followers forms a Nash equilibrium among them with analyses on the upper bound of the regret and suboptimality of performance.
In contrast to these studies, we focus on the situation where the follower is not myopic.

\citet{Zhang2020-hw} also considered a similar problem setting, where the reward setting is general-sum and the follower is not myopic.
The differences are that the transition function is unknown (i.e., in the RL setting), the follower's policy takes the leader's action as input, and only deterministic stationary policies are considered for the leader's policies.
The algorithm proposed by Zhang et al. (2020) shares its principal foundation with that of Bucarey et al. (2022) (see \cref{apdx:Zhang}).
This means that they use the same operator, implying that the limitation of the method of Bucarey et al. (2022) applies to that of Zhang et al. (2020).

Policy gradient methods have been proposed for SSGs with non-myopic followers \citep{Zheng2022-xj,vu2022stackelberg} based on the implicit function theorem \citep{Fiez2020-lu}.
They guarantee the convergence to differential SEs (DSEs), which are subsets of local SEs.
However, these methods are not applicable in our setting, except when using the surrogate model of the follower, due to their centralized nature of learning.
Moreover, the leader's policy obtained by these algorithms does not necessarily form the DSEs under the best-response follower because the follower's strategies in DSEs are not always the best response.

Several studies concentrate on the cooperative setting.
\citet{Kao2022-va} and \citet{Zhao2023-jc} analyzed the cooperative tasks with information asymmetry, where the leader cannot observe the follower's action~\citep{Kao2022-va} and the reward functions are known only to the follower~\citep{Zhao2023-jc}.
\citet{Kononen2004-xe} and \citet{Zhang2020-hw} point out Pareto efficiency and uniqueness as advantages of Stackelberg equilibria to Nash equilibria in cooperative tasks.

We proposed a novel algorithm for general-sum SSGs with the non-myopic best-response follower under the assumptions that the transition function is known and that the follower's best response is computable in a reasonable time.
Unlike the previous methods, iterative methods by the operator \citep{Bucarey2022-ro,Zhang2020-hw} and policy gradient methods \citep{Zheng2022-xj,vu2022stackelberg}, our algorithm guarantees monotone improvement of the leader's state values under the best-response follower.
Furthermore, we give a convergence guarantee even when no SSE policy exists. 

\section{Analysis on DP Operators}\label{sec:BucareyMethod}

Dynamic programming operators are the core of the solutions of single-agent MDPs and the foundation of RL algorithms \citep{Sutton2018-jf}.
For stochastic games, the operator is extended as an operation of solving a \emph{one-step game} by multiple players, and methods for Nash equilibria are proposed with convergence guarantees \citep{Hu2003-vy}.

For SSGs, however, existing methods based on such a one-step game operator \citep{Bucarey2022-ro,Zhang2020-hw} have problems in terms of the leader's performance in our setting.
This section describes it in detail.
In summary, we analyze the fixed points of the operator and show that, while the equilibrium formed by the fixed point holds under the best-response follower, the obtained leader's policy may perform poorly when the fixed point is not the SSE, since monotone improvement in the leader’s performance is not guaranteed.

The \emph{one-step game} operator $T:\mathcal{F}_{\mathcal{S}}\x\mathcal{F}_{\mathcal{S}}\to\mathcal{F}_{\mathcal{S}}\x\mathcal{F}_{\mathcal{S}}$ is defined for SSGs as 
\begin{multline}\label{eq:quasi-SEeq_Bucarey}
	(Tv)_i(s):=r_i(s,R_A(s,v),R_B(s,R_A(s,v),v_B)) \\+ \gamma_i\E_{s'\sim p(s'|s,R_A(s,v),R_B(s,R_A(s,v),v_B))}\left[v_i(s')\right]\quad(i\in\{A,B\}),
\end{multline}
where $v:=(v_A,v_B)\in \mathcal{F}_{\mathcal{S}}\times\mathcal{F}_{\mathcal{S}}$,
\begin{equation*}
	R_A(s,v):=\argmax_{f_s\in\Delta(\mathcal{A})} r_A(s,f_s,R_B(s,f_s,v_B)) + \gamma_A\E_{s'\sim p(s'|s,f_s,R_B(s,f_s,v_B))}\left[v_A(s')\right],
\end{equation*}
which is assumed to exist for all $s$ and $v$,%
\footnote{This assumption is necessary to derive a policy corresponding to a value function, and \citet{Bucarey2022-ro} implicitly assumes it. 
However, it is not guaranteed in general because the inside of $\argmax$ is not necessarily continuous with respect to $f_s$, even though $\Delta(\mathcal{A})$ is compact. 
Conversely, $R_B$ always exists because $\mathcal{B}$ is finite.} 
and
\begin{equation*}
	R_B(s,f_s,v_B):=\argmax_{b\in\mathcal{B}} r_B(s,f_s,b) \\+ \gamma_B\E_{s'\sim p(s'|s,f_s,b)}\left[v_B(s')\right],
\end{equation*}
which is assumed to be a singleton for all $s$, $f_s$, and $v_B$.%
\footnote{\citet{Bucarey2022-ro} defines the \emph{strong} version of $R_B$ similarly to the strong SSE. In this paper, we assume that $R_B$ is a singleton to simplify the discussion, in accordance with our goal of finding the SSE policies. 
As the core of our analysis does not depend on the uniqueness of $R_B$, the result can be easily extended to the method of Bucarey et al. (2022).}
\Cref{eq:quasi-SEeq_Bucarey} is regarded as a simplified version of the operator defined in \citet{Bucarey2022-ro}.
The single update of current values $v$ with $T$ as $v_i'(s)=(Tv)_i(s)$ can be seen as solving a normal-form (i.e., one-step) Stackelberg game for player $i$ at given state $s\in\mathcal{S}$ where the payoff functions are given by current Q functions \[q_i(s,f_s,b):=r_i(s,f_s,b)+\gamma_i\E_{s'\sim p(s'|s,f_s,b)}\left[v_i(s')\right]\quad(i\in\{A,B\})\] for the leader's mixed strategy $f_s$ and the follower's pure strategy $b$.
In this context, $R_B(s,f_s,v_B)$ is the follower's best response to $f_s$, $R_A(s,v)$ is the leader's Stackelberg equilibrium strategy, the pair $(R_A(s,v), R_B(s, R_A(s,v),v_B))$ is the Stackelberg equilibrium, and $((Tv)_A(s),(Tv)_B(s))$ is the equilibrium payoffs of the normal-form game for given $s$ and $v$.

\citet{Bucarey2022-ro} proposed a method to construct a stationary equilibrium policy corresponding to the fixed point of the operator, which is called \emph{fixed point equilibrium} (\emph{FPE}).
If there exists a fixed point $V:=(V_A,V_B)$ of $T$, the pair of the value functions of the FPE $(\bar{f},\bar{g})$ defined as, for any $s \in \mathcal{S}$,
\begin{align*}
	\bar{f}(s)\in R_A(s,V),
	\quad
	\bar{g}(s)\in R_B(s,\bar{f}(s),V_B)
\end{align*}
coincides with $V$~\citep{Bucarey2022-ro}.
Since $T$ is a contraction mapping for $v_A$ under certain conditions~\citep{Bucarey2022-ro}\footnote{One example is a condition that $r_A$ is an affine transformation of $r_B$ (i.e., cooperative or competitive with $r_B$).
In general, $T$ does not always converge, which is empirically demonstrated for the aforementioned example game (\Cref{fig:toytask}) in \citet{Jean-Marie2022-td}.}, we can obtain the fixed point asymptotically by repeatedly applying $T$ with arbitrary initial values.

Let us consider the characteristics of the fixed point $(V_A,V_B)$ of the operator $T$ in general situations where the fixed point exists.
\Cref{thrm:thrm_Bucarey} reveals a general property of $(V_A,V_B)$.
The proof is in \cref{apdx:thrm_Bucarey}.
(Appendices are provided as supplementary material.)

\begin{theorem}\label{thrm:thrm_Bucarey}
	For $V:=(V_A,V_B)\in\mathcal{F}_{\mathcal{S}}\x\mathcal{F}_{\mathcal{S}}$, let $R_A(V)\in\mathcal{W}_A$ be a stationary policy whose action distribution on a state $s\in\mathcal{S}$ is in $R_A(s,V)$ and $R_B(f,V_B)$ be a deterministic stationary policy under $f\in\mathcal{W}_A$ whose action on a state $s$ is in $R_B(s,f(s),V_B)$.
	Then, if $(TV)_i(s)=V_i(s)$ holds for all $i\in\{A,B\}$ and for all $s\in\mathcal{S}$, it holds that
	\begin{align*}
		V_A(s)=\max_{f\in\mathcal{W}_A}V_A^{fR_B(f,V_B)}(s)\enspace\forall s\in\mathcal{S};\quad
		V_B(s)=\max_{g\in\mathcal{W}_B}V_B^{R_A(V)g}(s)\enspace\forall s\in\mathcal{S}.
	\end{align*}
\end{theorem}
If $V$ is the fixed point, because $\bar{f}=R_A(V)$ and $\bar{g}=R_B(\bar{f},V_B)$ , it holds that
\begin{align}
	&V_B^{\bar{f}\bar{g}}(s)=V_B(s)=\max_{g\in\mathcal{W}_B}V_B^{\bar{f}g}(s),\label{eq:Bucarey_FPE_value_B}\\
	&V_A^{\bar{f}\bar{g}}(s)=V_A(s)=\max_{f\in\mathcal{W}_A}V_A^{fR_B(f,V_B)}(s)\label{eq:Bucarey_FPE_value_A}
\end{align}
for all $s\in\mathcal{S}$ in light of \cref{thrm:thrm_Bucarey}.
\Cref{eq:Bucarey_FPE_value_B} shows that the follower's FPE $\bar{g}=R_B(\bar{f},V_B)$ is the best response to the leader's FPE $\bar{f}$.
However, $R_B(f,V_B)$ is not the best response to arbitrary $f\in\mathcal{W}_A$, except when $V_B(s)=\max_{g\in\mathcal{W}_B}V_B^{fg}(s)~\forall s\in\mathcal{S}$.

The results of \cref{thrm:thrm_Bucarey} illuminate that the leader's performance $V_A$ of FPEs is the maximum of the leader's state value on $f\in\mathcal{W}_A$ under $R_B(f,V_B)$, but not the maximum under the follower's best responses $R_B^*(f)$ like SE value functions.
Therefore, the leader's performance of FPEs at each state can be less than the SE value, which is empirically demonstrated in \citet{Bucarey2019-bf}.
In addition, it is difficult to determine any meaning from $R_B(f,V_B)$ for arbitrary $f\in\mathcal{W}_A$, which makes the performance of FPEs unpredictable.

Although \citet{Bucarey2022-ro} derives the sufficient condition for SSEs to exist and for FPEs to coincide with the SSEs, this condition is not compatible with our problem setting or can be a strong assumption.
The sufficient condition is equivalent to the union of two conditions \citep{Bucarey2022-ro}.
The first is that the follower's discount rate $\gamma_B=0$, which is a condition on the follower's learning algorithm, but we admit no other assumptions on the follower than the best responsivity.
The second is that the transition function does not depend on the follower's actions, which can limit the applications of their approach.
In the end, finding FPEs does not guarantee the leader's performance in our setting.

To overcome the problem of SSGs that SSEs do not always exist, in the next section, we propose an algorithm that guarantees monotone improvement in the leader's performance toward alternative equilibria, \emph{Pareto-optimal policies}.

\section{Pareto-Optimal Policy Iteration}\label{sec:proposition}

SSE policies do not always exist in general-sum stochastic games.
Therefore, it is desired that an algorithm converges to an SSE policy if it exists and to a policy that is ``reasonable'' provided an SSE policy does not exist. First, we introduce the notion of \emph{Pareto-optimality} as the reasonable target. A policy that approximates a Pareto-optimal value function with an arbitrary precision always exists, and it admits an SSE policy if an SSE policy exists. 
Then, we derive a \emph{policy improvement theorem} for SSGs under best-response followers. 
Based on the policy improvement theorem, we design an algorithm that monotonically improves the policy toward Pareto-optimal policies.

\paragraph{Notation}
We introduce the notation for the dominance relation. 
Let $v, v' \in \mathcal{F}_\mathcal{S}$. 
If $v(s) = v'(s)$ for all $s \in \mathcal{S}$, we express $v \eqs v'$.
If $v(s) \ge v'(s)$  for all $s \in \mathcal{S}$, then $v$ weakly dominates $v'$ and $v \succcurlyeq v'$. 
If $v \succcurlyeq v'$ and $v \not\eqs v'$, $v$ strictly dominates $v'$ and $v \succ v'$.

\subsection{Pareto-Optimal Policy}\label{sec:po_policy}

SSE policies exist if and only if there exists $f\in\mathcal{W}_A$ that maximizes $V_A^{f\dag}(s)$ for all $s\in\mathcal{S}$.
We consider $V_A^{f\dag}(s)$ as $|\mathcal{S}|$ objective functions with $f\in\mathcal{W}_A$ as a common input.
From this multi-objective optimization viewpoint, we define the Pareto-optimal policies.
Hereafter, we view a value function $v$ as a vector in $\R^{\abs{\mathcal{S}}}$ and apply topological argument in the standard sense in $\R^{\abs{\mathcal{S}}}$.

\begin{definition}[Pareto Optimality]\label{thrm:def_POP}
	Let $\mathcal{V} = \{v \in \mathcal{F}_\mathcal{S} : v \eqs V_A^{f\dag}\ \exists f \in \mathcal{W}_A\}$ be the set of a reachable value function. 
	Let $\partial \mathcal{V}$ be the boundary of $\mathcal{V}$ and 
	$\mathrm{cl}\mathcal{V} = \mathcal{V} \cup \partial\mathcal{V}$ be the closure of $\mathcal{V}$. 
	A Pareto-optimal (PO) value function $v^* \in \mathrm{cl}\mathcal{V}$ is such that there exists no $v \in \mathrm{cl}\mathcal{V}$ satisfying $v \succ v^*$. The set of PO value functions is denoted as $\mathcal{PV}$.
	A stationary policy $f\in \mathcal{W}_A$ satisfying $V_A^{f\dag} \in \mathcal{PV}$ is a PO policy. 
\end{definition}

There always exists $\mathcal{PV} \neq \emptyset$. Moreover, $\mathcal{PV} \subseteq \partial\mathcal{V}$. 
A PO policy exists if and only if $\mathcal{PV} \cap \mathcal{V} \neq \emptyset$. 
For any PO value function $v^* \in \mathcal{PV}$, there always exists a policy whose value function is arbitrarily close to $v^*$. 
Meanwhile, there is no stationary policy that realizes a value function better than a PO value function.
Moreover, an arbitrary PO policy is an SSE policy if there exist SSE policies, as shown in \cref{thrm:thrm_SEimpliesPO}.
Therefore, PO value functions are a reasonable alternative to the SE value function.
The proof is presented in \cref{apdx:thrm_SEimpliesPO}.

\begin{proposition}\label{thrm:thrm_SEimpliesPO}
	The SE value function $V_A^* \in \mathcal{PV}$ if and only if $\mathcal{PV}$ is a singleton. 
	Consequently, the SSE policies exist if and only if $\mathcal{PV}$ is a singleton and $\mathcal{PV} \subseteq \mathcal{V}$.
\end{proposition}

\begin{figure}[t]
\centering
\includegraphics[width=0.45\hsize,clip,trim=0 0 0 0]{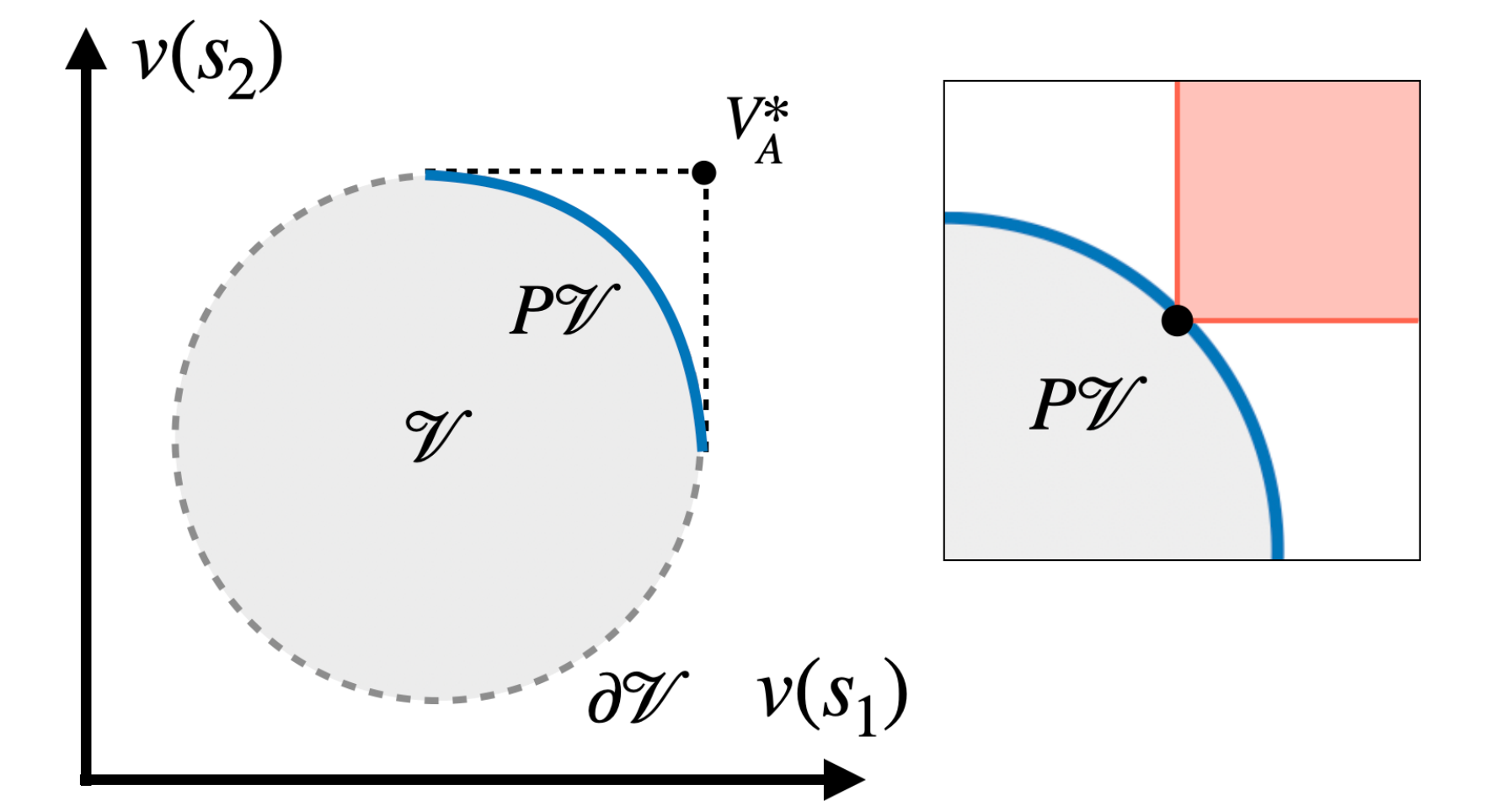}%
\caption{Pareto optimal value functions in the two-state value function space. Thick blue line: Pareto optimal value functions. Red shaded area (the upper right of the dot in the box): the area of value functions dominating the value function represented by the black dot, which is empty for PO value functions.}
\label{fig:PO}
\end{figure}

\subsection{Policy Improvement Theorem}\label{sec:improvement}

We derive three important lemmas based on which we design the proposed approach.

\Cref{thrm:lemma_improve} is an extension of the policy improvement theorem~\citep{Sutton2018-jf} in single-agent MDPs.
With this theorem, we can confirm if a policy $f$ is better than a baseline $f'$ without computing the value function for $f$. 
The proof is shown in \cref{apdx:lemma_improve}.
The key idea behind its proof is to express the follower's best response in terms of the leader's policy and then apply reasoning analogous to the policy improvement theorem for single-agent MDPs.

\begin{theorem}\label{thrm:lemma_improve}
	Let $f\in\mathcal{W}_A,f'\in\mathcal{W}_A$ be stationary policies, and a Q-function $Q_A^{f'\dag}(s,f): \mathcal{S}\x\mathcal{W}_A\to\R$ is expressed as
	\begin{equation*}
		Q_A^{f'\dag}(s,f):=r_A(s,f(s),R_B^*(s,f))+\gamma_A\E_{s'}\left[V_A^{f'\dag}(s')\right],
	\end{equation*}
	where $\E_{s'}$ is considered under $p(s'|s,f(s),R_B^*(s,f))$. 
	Then, the following conditions hold:
        \begin{align*}
            &(a)\quad Q_A^{f'\dag}(\cdot ,f) \succcurlyeq V_A^{f'\dag}\implies V_A^{f\dag} \succcurlyeq Q_A^{f'\dag}(\cdot,f)\succcurlyeq V_A^{f'\dag};\\
            &(b)\quad Q_A^{f'\dag}(\cdot,f)\eqs V_A^{f'\dag} \iff V_A^{f\dag}\eqs V_A^{f'\dag};\\
            &(c)\quad Q_A^{f'\dag}(\cdot ,f) \succ V_A^{f'\dag} \iff V_A^{f\dag} - V_A^{f'\dag} \succ \gamma_A\E_{s'}\left[V_A^{f\dag}(s') - V_A^{f'\dag}(s')\right].
        \end{align*}
\end{theorem}

A necessary condition for PO policies is given below, which shows that no further performance improvement can be made by \cref{thrm:lemma_improve}~(a) from PO policies. 
This proof is in \cref{apdx:lemma_POimplies}.

\begin{theorem}[Necessary condition for PO policies]\label{thrm:lemma_POimplies}
	If $f'\in \mathcal{W}_A$ is a PO policy, it holds that for any $f\in \mathcal{W}_A$
	\begin{equation}
		Q_A^{f'\dag}(\cdot,f)\succcurlyeq V_A^{f'\dag}
		\implies V_A^{f\dag}\eqs V_A^{f'\dag}.\label{eq:lemma_POimplies}
	\end{equation}
\end{theorem}

The following result shows a sufficient condition for PO policies.
This proof is presented in \cref{apdx:lemma_impliesPO}.

\begin{theorem}[Sufficient condition for PO policies]\label{thrm:lemma_impliesPO}
	$f'\in \mathcal{W}_A$ is a PO policy if, for any $f\in \mathcal{W}_A$, either of the following two conditions hold:
        \begin{align*}
            &(\mathrm{i})\quad Q_A^{f'\dag}(\cdot,f) \eqs V_A^{f'\dag};\\
            &(\mathrm{ii})\quad \exists s \in \mathcal{S}, Q_A^{f'\dag}(s,f)<V_A^{f'\dag}(s)-\dfrac{\gamma_A}{1-\gamma_A}\delta(f,f'),
        \end{align*}
        where $\delta(f,f'):=\max\limits_{s\in\mathcal{S}}\E_{s' \sim p(\cdot|s,f(s),R_B^*(s,f))}\left[ Q_A^{f'\dag}(s',f)-V_A^{f'\dag}(s') \right].$
\end{theorem}
When $\gamma_A=0$, \cref{thrm:lemma_impliesPO} implies that no further policy update by \cref{thrm:lemma_improve}~(a) with strict improvement is a sufficient condition for PO policies.
Combined with \cref{thrm:lemma_POimplies}, $f'$ is a PO policy if and only if there is no room for policy improvement when $\gamma_A=0$.

\subsection{Policy Iteration for PO Policies}\label{sec:approach}

We propose an approach that monotonically improves policy toward PO policies and is guaranteed to converge. 

PO value functions are not always unique. 
To guide the algorithm, we introduce scalarization.
Given a stationary policy $f\in\mathcal{W}_A$, 
we maximize a Pareto-compliant scalarization $\mathcal{L}$ of the state value on $f$, namely, $\mathcal{L}[  V_A^{f\dag}]$, where the Pareto-compliant scalarization\footnote{
	For example, a weighted sum scalarization $\mathcal{L}[v] = \sum_{s \in \mathcal{S}} \alpha_s v(s)$ is a Pareto-compliant scalarization for $\boldsymbol{\alpha}:=\{\alpha_s\in\R_{>0}\}_{s\in\mathcal{S}}$ such that $\sum_{s\in\mathcal{S}}\alpha_s=1$.
	If we choose $\alpha$ such that $\alpha_s = \rho(s)$ for all $s \in \mathcal{S}$, then $\mathcal{L}[v]$ coincides with the expected return under the initial state distribution $\rho$.
} $\mathcal{L}: \mathrm{cl}\mathcal{V} \to \R$ is a continuous function such that $v \succ v' \implies \mathcal{L}[v] > \mathcal{L}[v']$. 
Then, the maximum of the scalarized value in $\mathrm{cl}\mathcal{V}$ is obtained by a PO value function, namely,
$\argmax_{v \in \mathrm{cl}\mathcal{V}} \mathcal{L}\left[ v \right] \subseteq \mathcal{PV}$.

Directly maximizing $\mathcal{L}[V_A^{f\dag}]$ is, however, intractable.
This is because the computation of $V_A^{f\dag}$ for each $f \in \mathcal{W}_A$ requires
repeatedly applying the Bellman expectation operator under $f$ and $R_B^*(f)$,
\begin{equation*}
	(T_A^{f R_B^*(f)}V)(s):=r_A(s,f(s),R_B^*(s,f))\\+\gamma_A\E_{s'\sim p(s'|s,f(s),R_B^*(s,f))}\left[V(s')\right],
\end{equation*}
to an initial function $V_0\in\mathcal{F}_{\mathcal{S}}$ until it converges, where the convergence is guaranteed as in a single-agent MDP.

To alleviate this difficulty and obtain a sequence $\{f_t\}_{t\ge 0}^{\infty}$ satisfying $\mathcal{L}[V_A^{f_{t+1}\dag}] \ge \mathcal{L}[V_A^{f_{t}\dag}]$, we employ the policy improvement theorem, as in the policy iteration in single-agent MDPs. Rather than computing $V_A^{f\dag}$ for each $f$, we calculate $Q_A^{f_t\dag}(\cdot, f)$, requiring only $V_A^{f_t\dag}$. Then, we construct the set $\mathcal{W}_\succcurlyeq(f_t)$ of policies satisfying the policy improvement condition from $f_t$, namely,
\begin{equation}    \mathcal{W}_\succcurlyeq(f_t):=\left\{f\in\mathcal{W}_A~|~Q_A^{f_t\dag}(\cdot,f)\succcurlyeq V_A^{f_t\dag}\right\}.\label{eq:def_hat_W}
\end{equation}
Selecting the next policy $f_{t+1}$ from $\mathcal{W}_\succcurlyeq(f_t)$ guarantees $\mathcal{L}[V_A^{f_{t+1}\dag}] \ge \mathcal{L}[V_A^{f_{t}\dag}]$ in light of \cref{thrm:lemma_improve}~(a).
Ideally,\footnote{
	In practice, 
	$\argmax$ is not necessary, and sometimes it 
	does not exist or is not computable. 
	An alternative  approach is to select a policy $f_{t+1}$ from a set $\mathcal{W}_{1-\epsilon}(f_t)$ of $\epsilon$-optimal policies in terms of $\mathcal{L}$, such as
	\begin{multline}
		\mathcal{W}_{1-\epsilon}(f_t) := \bigg\{ f \in \mathcal{W}_\succcurlyeq(f_t) \mid
		\mathcal{L}\left[ Q_A^{f_t\dag}(\cdot,f) \right] - \mathcal{L}\left[ V_A^{f_t\dag}\right] \\
		\ge (1 - \epsilon) \sup_{f \in \mathcal{W}_\succcurlyeq(f_t)} \bigg( \mathcal{L}\left[ Q_A^{f_t\dag}(\cdot,f) \right] - \mathcal{L}\left[ V_A^{f_t\dag}\right] \bigg) \bigg\}\label{eq:Wcond}
	\end{multline}
	for some $\epsilon \in [0, 1)$. 
	Notably, $\mathcal{W}(f_t) = \mathcal{W}_{1}(f_t)$. Therefore, the update in \cref{eq:update} is a special case of \cref{eq:Wcond}.
	If it is still intractable, selecting $f_{t+1}\in\mathcal{W}_\succcurlyeq(f_t)$ s.t. $Q_A^{f_t\dag}(\cdot,f_{t+1})\succ V_A^{f_t\dag}$ ensures that (a) and (b) in \cref{thrm:thrm_algo} hold.} the proposed algorithm selects $f_{t+1}$ from the set of policies that maximizes the scalarized value, namely,
\begin{align}
	f_{t+1}\in\mathcal{W}(f_t):=\argmax_{f\in\mathcal{W}_\succcurlyeq(f_t)} \mathcal{L}\left[ Q_A^{f_t\dag}(\cdot,f) \right].\label{eq:update}
\end{align}

\Cref{thrm:thrm_algo} guarantees that the proposed algorithm always converges and that the limit satisfies the necessary condition for PO policies (i.e., $v_\infty \in \partial \mathcal{V}$).
The proof is shown in \cref{apdx:thrm_algo}.

\begin{theorem}\label{thrm:thrm_algo}
	Let $f_0\in\mathcal{W}_A$ be an arbitrary initial policy and a policy sequence $\{f_{t}\in\mathcal{W}(f_{t-1})\}_{t=1}^{\infty}$ be obtained by \cref{eq:Wcond}.
	Let $\{v_t = V_A^{f_t\dag}\}_{t=0}^{\infty}$ be the corresponding value functions.
	Then, the following statements hold:
	\begin{itemize}\setlength\itemsep{0em}
		\item[(a)] $\{v_t\}_{t=0}^{\infty}$ monotonically increases in the sense that $v_{t+1} \succcurlyeq v_t$ and converges to $v_\infty \eqs \lim_{t\to\infty} v_t$;
		\item[(b)] $v_{t+1} \eqs v_t$ if and only if $v_t \eqs v_\infty$;
		\item[(c)] $v_\infty \in \partial \mathcal{V}$ and $\min_{s \in \mathcal{S}} v(s) - v_\infty(s)  \le \gamma_A\left( \max_{s \in \mathcal{S}} v(s) - v_\infty(s) \right)$ for all $v\in \mathcal{V}$;
		\item[(d)] $v_\infty \in \mathcal{PV}$ if $\gamma_A = 0$.
	\end{itemize}
\end{theorem}

As shown in the proof (\cref{apdx:thrm_algo}), the algorithm ceases to improve if and only if $Q_A^{f_t\dag}(\cdot,f) \eqs V_A^{f_t\dag}$ for all $f\in \mathcal{W}_\succcurlyeq(f_t)$.
The algorithm terminates when it holds and returns $f_t$.
The entire proposed algorithm is shown in \cref{algo:proposed}.

\begin{algorithm}[t]
	\caption{Pareto-Optimal Policy Iteration}
	\label{algo:proposed}
	\begin{algorithmic}[1]
		\REQUIRE{Maximum number of iterations $M$.}
		\STATE {Randomly initialize $f_0\in\mathcal{W}_A$.}
		\FOR {$t=0$ to $M-1$}
		\STATE {Compute $V_A^{f_t\dag}$ by repeatedly applying $T_A^{f_tR_B^*(f_t)}$}
		\STATE {Compute $\mathcal{W}_\succcurlyeq(f_t)$ in \cref{eq:def_hat_W}}
		\IF {$Q_A^{f_t\dag}(s,f) \eqs V_A^{f_t\dag}$ for all $f\in\mathcal{W}_\succcurlyeq(f_t)$} \label{algo:proposed:1_l6}
		\STATE {{\bf return} $f^*\leftarrow f_t$}
		\ELSE
		\STATE{Compute $\mathcal{W}(f_t)$ in \cref{eq:update} or \cref{eq:Wcond}}
		\STATE {Randomly sample $f_{t+1}$ from $\mathcal{W}(f_t)$}
		\ENDIF
		\ENDFOR
		\STATE {{\bf return} $f^*\leftarrow f_M$}
		\ENSURE{A stationary policy $f^*$}
	\end{algorithmic}
\end{algorithm}

\begin{figure}[t]
\centering
\includegraphics[width=0.5\hsize,clip,trim=0 0 0 0]{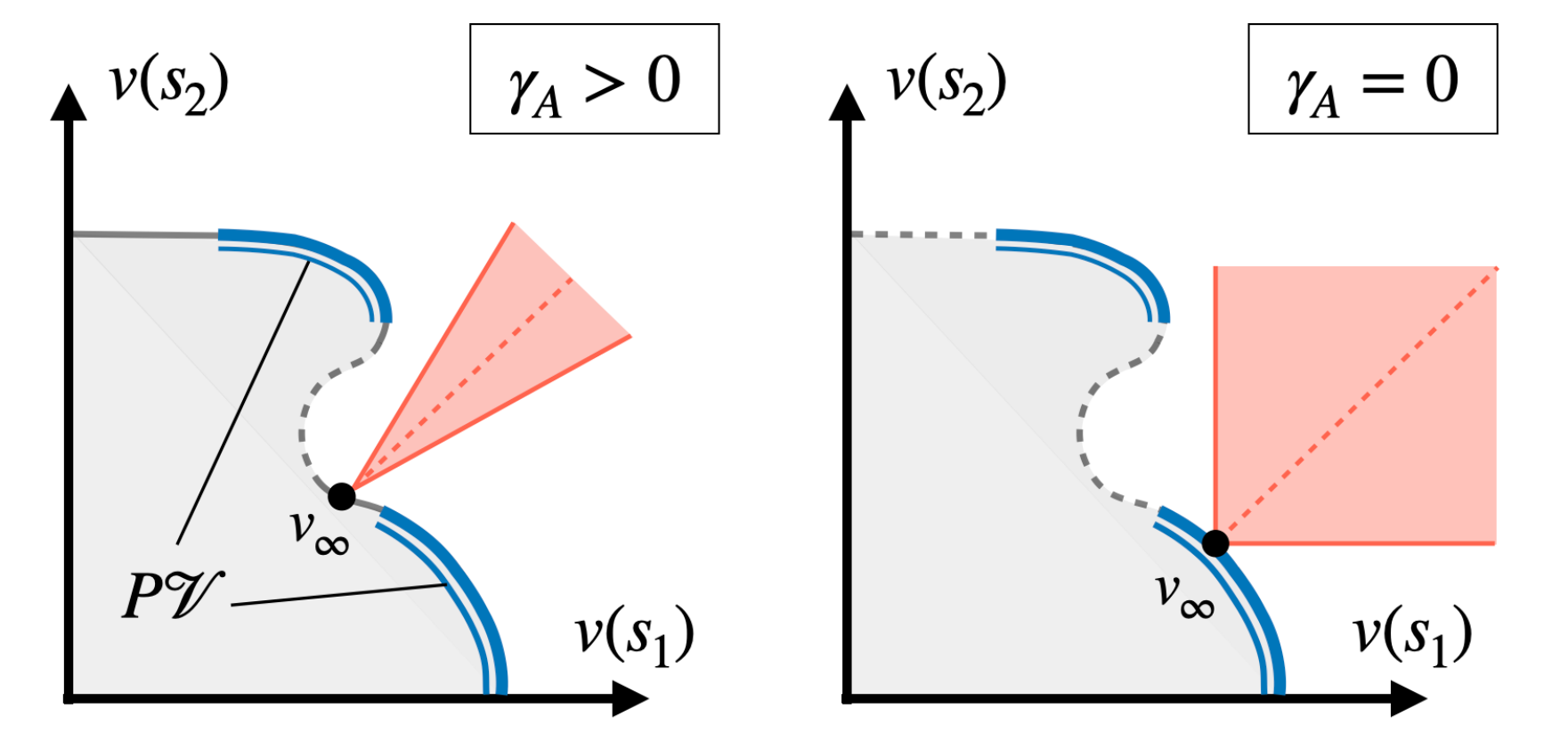}%
\caption{Visualization of \cref{thrm:thrm_algo} (c) and (d). 
For $v_\infty$, there must not exist a value function in the red-shaded cone 
representing the area of $v$ violating the inequality in (c).
The upper and lower slopes of the cone are $1/\gamma_A$ and $\gamma_A$, respectively.
$v_{\infty}$ can be on the non-Pareto boundary (grey solid line) when $\gamma_A>0$ (left), but $v_{\infty}$ is on $\mathcal{PV}$ when $\gamma_A=0$ (right) because the cone equals the area of dominating value functions.}
\label{fig:theorem6_6}
\end{figure}

The convergence guarantee and the monotone improvement of \cref{algo:proposed} (Statements (a) and (b) in \cref{thrm:thrm_algo}) show the advantages of the proposed approach.
However, compared with the policy iteration approach in single-agent MDPs, where the convergence to the optimal policy is guaranteed, the fixed point of \cref{algo:proposed} is not assured to be PO value functions. Rather, \cref{thrm:thrm_algo}~(c) guarantees that the limit is at the boundary $\partial \mathcal{V}$, including all the PO value functions.
In a special case of $\gamma_A = 0$, the limit is guaranteed to be a PO value function (\cref{thrm:thrm_algo}~(d)).
The visualization of (c) and (d) is in \cref{fig:theorem6_6}.

This condition, the leader's discount rate $\gamma_A=0$, is a condition on the leader's learning process, meaning \cref{algo:proposed} does not need any assumptions on the follower's learning process to guarantee the convergence and the leader's ``reasonable'' performance.
This property allows us to apply this algorithm to various applications modeled by decentralized SSGs in which we are the leader.
We give an example of such applications in \cref{sec:application}.

There is another algorithmic difference from the policy iteration for the single-agent MDP.
In single-agent MDPs, a state-wise maximization of the action value function of $f_t$,
$a_s \in \argmax_{a \in \mathcal{A}} Q^{f_t}(s, a)$ is performed to construct the next (deterministic) policy.
This is invalid in SSGs because such a policy does not necessarily satisfy the condition of \cref{thrm:lemma_improve}~(a) and may not improve the values. 
Moreover, the optimal policy is not necessarily a deterministic one.
Because of these differences, \cref{eq:update} cannot be simplified to the state-wise maximization. 
This can be a limitation from a practical viewpoint when \cref{eq:update} (or \cref{eq:Wcond}) is intractable due to the large search space of $\mathcal{W}_A$.
We propose a practical strategy of splitting the policy space to find the next policy efficiently, avoiding exhaustive search over the entire space $\mathcal{W}_A$.
Its details are provided in \cref{sec:improve_algo_feasibility}

\section{Limitation and Future Work}\label{sec:conclusion}



This study proposed a novel algorithm with a convergence guarantee in two-player general-sum SSGs under best-response followers.
We introduced the notion of the Pareto-optimal value function to target it even if there are no SSEs, and developed an algorithm that monotonically improves the policy toward the Pareto front.
While existing methods have little guarantee under best-response followers, especially in games where the SSEs do not exist, our proposed approach can monotonically improve the leader's performance and guarantees its limit to be Pareto optimal in state values when the leader is myopic.
To the best of our knowledge, this is the first theoretical guarantee in general-sum myopic-leader SSGs.

However, there is room for improvement both from theoretical and practical viewpoints.
Further research must focus on addressing the limitations of the current work listed below. 
First, the Pareto-optimality of the proposed algorithm is not guaranteed. A necessary and sufficient condition for the Pareto-optimality of the algorithm must be derived. The development of a restart strategy to satisfy such a sufficient condition is desired.
Second, a computationally efficient update rule to select the next policy (cf. \cref{eq:update} or \cref{eq:Wcond}) is required. Practically, a state-wise update similar to the policy iteration in single-agent MDPs is desired. 
Finally, our approach requires knowledge of the follower's best response and transition probability, as in the previous studies. A sample approximation of the algorithm by reinforcement learning is desired to widen the applicability of the proposed approach.

\newpage
\appendix

\section{Proof of \texorpdfstring{\cref{thrm:thrm_Bucarey}}{Lg}}\label{apdx:thrm_Bucarey}

\begin{proof}
	First, we prove that $(TV)_A \eqs V_A$ implies $V_A \eqs \max_{f\in\mathcal{W}_A} V_A^{fR_B(f,V_B)}$ for a fixed $V_B$.
	
	Suppose that $V_A \eqs (TV)_A$.
	Given a policy $f \in \mathcal{W}_A$, let $f_s = f(s)$ for each $s \in \mathcal{S}$.
	Then, we obtain for all $s \in \mathcal{S}$
	\begin{subequations}
		\begin{align}
			V_A(s) &= (TV)_A(s) \\
			&= \sup_{f_s \in \Delta(\mathcal{A})} r_A(s,f_s,R_B(s,f_s,V_B))+\gamma_A\E_{s'\sim p(s'|s,f_s,R_B(s,f_s,V_B))}\left[V_A(s')\right] \\
			&\ge r_A(s,f_s,R_B(s,f_s,V_B))+\gamma_A\E_{s'\sim p(s'|s,f_s,R_B(s,f_s,V_B))}\left[V_A(s')\right] \\
			&= r_A(s,f(s),R_B(s,f(s),V_B))+\gamma_A\E_{s'\sim p(s'|s,f(s),R_B(s,f(s),V_B))}\left[V_A(s')\right].
		\end{align}
	\end{subequations}
	By repeatedly applying this inequality, we obtain
	\begin{subequations}\label{eq:V_Age_subeq}
		\begin{align}
			V_A(s)
			&\ge r_A(s,f(s),R_B(s,f(s),V_B))+\gamma_A\E_{s_1}\left[V_A(s_1)\right]\\
			&=\begin{multlined}[t]
				r_A(s,f(s),R_B(s,f(s),V_B))\\+\gamma_A\E_{s_1}\Bigl[
				r_A(s_1,f(s_1),R_B(s_1,f(s_1),V_B))+\gamma_A\E_{s_2}\left[V_A(s_2)\right]\Bigl]
			\end{multlined}\\
			&\ge \cdots\nonumber\\
			&=\begin{multlined}[t]
			    \E^{f R_B(f,V_B)}\left[\sum_{t=0}^{n-1}\gamma_A^t\hat{r}_A(s_t,a_t,b_t)\middle| s_0=s\right] 
                    + \gamma_A^n \E^{f R_B(f,V_B)}\Bigl[V_A(s_n)\Bigl| s_0=s\Bigl].
			\end{multlined}\label{eq:V_Age_recc}
		\end{align}
	\end{subequations}
	In the limit for $n\to\infty$, the first term of the right-hand side of \cref{eq:V_Age_recc} converges to $V_A^{f R_B(f,V_B)}(s)$, while the second term converges to 0\footnote{Since $\hat{r}_A$ is bounded and $\gamma_A\in[0,1)$, the inside of the expectation of \cref{eq:V_Age_recc} is always bounded, which enables the exchange of the extreme and the expectation. Therefore, the same exchanges are performed in \cref{eq:V_Bge_recc} and \cref{eq:improve_infty}.}.
	As the choice of $f$ is arbitrary, we obtain $V_A \succcurlyeq \sup_{f \in \mathcal{W}_A} V_A^{f R_B(f,V_B)}$.
	
	We show $V_A \eqs \max_{f \in \mathcal{W}_A} V_A^{f R_B(f,V_B)}$.
	It suffices to show that there exists a policy $f^* \in \mathcal{W}_A$ such that $V_A \eqs V_A^{f^* R_B(f^*,V_B)}$.     
	Because $R_A(s, V)$ is non-empty, let $f_s^* \in R_A(s, V)$ for each $s \in \mathcal{S}$ and $f^*$ be a policy whose action distribution at $s$ is $f^*(s) = f_s^*$ for each $s$, we have for all $s \in \mathcal{S}$
	\begin{subequations}
		\begin{align}
			V_A(s) &= (TV)_A(s) \\
			&= \sup_{f_s \in \Delta(\mathcal{A})} r_A(s,f_s,R_B(s,f_s,V_B))+\gamma_A\E_{s'\sim p(s'|s,f_s,R_B(s,f_s,V_B))}\left[V_A(s')\right] \\
			&= r_A(s,f_s^*,R_B(s,f_s^*,V_B))+\gamma_A\E_{s'\sim p(s'|s,f_s^*,R_B(s,f_s^*,V_B))}\left[V_A(s')\right] \\
			&=\begin{multlined}[t]
			    r_A(s,f^*(s),R_B(s,f^*(s),V_B))
                    +\gamma_A\E_{s'\sim p(s'|s,f^*(s),R_B(s,f^*(s),V_B))}\left[V_A(s')\right].
			\end{multlined}
		\end{align}
	\end{subequations}
	Therefore, with the repeated application of this equality, analogously to the above argument, we obtain $V_A \eqs V_A^{f^* R_B(f^*,V_B)}$.
	
	Second, letting $V_A$ be fixed, we prove that $(TV)_B \eqs V_B$ implies $V_B \eqs \max_{g\in\mathcal{W}_B}V_B^{fg}$ for $f=R_A(V)$.
	
	Suppose $V_B \eqs (TV)_B$ holds.
	Then, by constructing $R_B$, we obtain
	\begin{align}
		V_B(s) = (TV)_B(s) \ge r_B(s,R_A(s,V),b_s)+\gamma_B\E_{s'\sim p(s'|s,R_A(s,V),b_s)}\left[V_B(s')\right]
	\end{align}
	for all $s\in\mathcal{S}$ and for all $b_s\in\mathcal{B}$.
	It implies that
	\begin{align}
		V_B(s)\ge r_B(s,R_A(s,V),g(s))+\gamma_B\E_{s'\sim p(s'|s,R_A(s,V),g(s))}\left[V_B(s')\right]\label{eq:V_Bge}
	\end{align}
	for all $s\in\mathcal{S}$ and for all $g\in\mathcal{W}_B^d$.
	By recursively applying \cref{eq:V_Bge} for $V_B$ on the right-hand side, we have for all $s\in\mathcal{S}$    \begin{subequations}\label{eq:V_Bge_subeq}
		\begin{align}
			V_B(s)&\ge r_A(s,R_A(s,V),g(s))+\gamma_B\E_{s_1}\left[V_B(s_1)\right]\\
			&\ge\begin{multlined}[t]
				r_A(s,R_A(s,V),g(s))+\gamma_B\E_{s_1}\Bigl[
				r_A(s_1,R_A(s_1,V),g(s_1))
                    +\gamma_B\E_{s_2}\left[V_B(s_2)\right]\Bigl]
			\end{multlined}\\
			&\ge \cdots\nonumber\\
			&\ge\begin{multlined}[t]
			    \E^{R_A(V)g}\left[\sum_{t=0}^{n-1}\gamma_B^t\hat{r}_B(s_t,a_t,b_t)\middle| s_0=s\right] 
                    +\gamma_B^n\E^{R_A(V)g}\Bigl[V_B(s_n)\Bigl| s_0=s\Bigl].
			\end{multlined}\label{eq:V_Bge_recc}
		\end{align}
	\end{subequations}
	In the limit for $n\to\infty$, the first term of the right-hand side of \cref{eq:V_Bge_recc} converges to $V_B^{R_A(V)g}(s)$, and the second term converges to 0. 
	Therefore, $V_B(s)\ge V_B^{R_A(V)g}(s)$ holds for all $s\in\mathcal{S}$.
	Since $g\in\mathcal{W}_B^d$ is arbitrary, it holds that
	\begin{align}
		V_B(s)\succcurlyeq\max_{g\in\mathcal{W}_B^d}V_B^{R_A(V)g}.
	\end{align}
	Owing to the existence of a deterministic optimal policy for any single-agent MDP, we obtain
	\begin{align}
		V_B \succcurlyeq\max_{g\in\mathcal{W}_B^d}V_B^{R_A(V)g}\eqs\max_{g\in\mathcal{W}_B}V_B^{R_A(V)g}.
	\end{align}
	
	We show that $V_B(s)\eqs \max_{g\in\mathcal{W}_B}V_B^{R_A(V)g}$.
	It suffices to show that there exists $g^*$ such that $V_B(s)\eqs V_B^{R_A(V)g^*}$.
	Let $g^*$ such that $g^*(s) = R_B(s,R_A(s,V),V_B)$ for each $s \in \mathcal{S}$. 
	Then, by the definition of $R_B$, for any $s \in \mathcal{S}$, we have
	\begin{align}
		V_B(s) = r_B(s,R_A(s,V),g^*(s))+\gamma_B\E_{s'\sim p(s'|s,R_A(s,V),g^*(s))}\left[V_B(s')\right].
	\end{align}
	With the repeated application of this equality, analogous to the above argument, we obtain $V_B \eqs V_B^{R_A(V)g^*}$.
\end{proof}

\section{Proof of \texorpdfstring{\cref{thrm:thrm_SEimpliesPO}}{Lg}}\label{apdx:thrm_SEimpliesPO}

\begin{proof}
	First, suppose that $V_A^* \in \mathcal{PV}$.
	Then, by the definition of the SE value function, we have $V_A^* \succcurlyeq v$ for all $v \in \mathrm{cl}\mathcal{V}$. 
	Therefore, no other PO value function exists. 
	Hence, $\mathcal{PV} = \{V_A^*\}$. 
	
	Subsequently, suppose that $\mathcal{PV} = \{v^*\}$ is a singleton.
	Then, $v^* \succ v$ for all $v \in \mathcal{V}\setminus \mathcal{PV}$ because,  for all $v\in\mathcal{V}\setminus \mathcal{PV}$, there exists $\hat{v}\in\mathcal{PV}$ such that $\hat{v}\succ v$. 
	Suppose that $v^* \not \eqs V_A^*$. Then, for some $s \in \mathcal{S}$, $v^*(s) < \sup_{f \in \mathcal{W}} V_A^{f\dag}(s)$ holds. It implies that there exists $v \in \mathcal{V}\setminus \mathcal{PV}$ such that $v^*(s) < v(s)$ for some $s$, which contradicts $v^* \succ v$. 
	Hence, $v^* \eqs V_A^*$.
	
	Altogether, we have $V_A^* \in \mathcal{PV}$ if and only if $\mathcal{PV}$ is a singleton.
	Because PO policies exists if and only if $\mathcal{PV} \cap \mathcal{V} \neq \emptyset$, the SE policy exists if and only if $\mathcal{PV} = \{v^*\}$ is a singleton and $v^* \in \mathcal{V}$.
\end{proof}

\section{Proof of \texorpdfstring{\cref{thrm:lemma_improve}}{Lg}}\label{apdx:lemma_improve}

\begin{proof}
	We prove each statement one by one.
	\paragraph{Proof of (a)}
	The definition of $Q_A^{f'\dag}(s,f)$ is equivalent to
	\begin{align}
		Q_A^{f'\dag}(s,f)=\E^{fR_B^*(f)}\left[\hat{r}_A(s,a,b)\middle|s\right]+\gamma_A\E^{fR_B^*(f)}\left[V_A^{f'\dag}(s')\middle|s\right].
	\end{align}
	Then, if it holds that $V_A^{f'\dag}(s)\le Q_A^{f'\dag}(s,f)$ for all $s\in\mathcal{S}$, we have, for all $s_0\in\mathcal{S}$,
	\begin{subequations}\label{eq:improve_proof}
		\begin{align}
			V_A^{f'\dag}(s_0)&\le\E^{fR_B^*(f)}\left[\hat{r}_A(s_0,a_0,b_0)\middle|s_0\right]+\gamma_A\E^{fR_B^*(f)}\left[V_A^{f'\dag}(s_1)\middle|s_0\right]\\
			&\le\begin{multlined}[t]
				\E^{fR_B^*(f)}\left[\hat{r}_A(s_0,a_0,b_0)\middle|s_0\right]+\gamma_A\E^{fR_B^*(f)}\Bigl[\\
				\E^{fR_B^*(f)}\left[\hat{r}_A(s_1,a_1,b_1)\middle|s_1\right]+\gamma_A\E^{fR_B^*(f)}\left[V_A^{f'\dag}(s_2)\middle|s_1\right]\Bigl|s_0\Bigl]
			\end{multlined}\\
			&\le\dots\nonumber\\
			&\le\E^{fR_B^*(f)}\left[\sum_{t=0}^{n-1}\gamma_A^t\hat{r}_A(s_t,a_t,b_t)\middle|s_0\right] + \gamma_A^n\E^{fR_B^*(f)}\left[V_A^{f'\dag}(s_n)\middle|s_0\right].\label{eq:improve_infty}
		\end{align}
	\end{subequations}
	In the limit for $n\to\infty$, the first term converges to $V_A^{fR_B^*(f)}(s_0)$ and the second term converges to 0. 
	Therefore, it holds that $V_A^{f'\dag}(s_0)\le Q_A^{f'\dag}(s,f)\le V_A^{fR_B^*(f)}(s_0)=V_A^{f\dag}(s_0)$ for all $s_0\in\mathcal{S}$.
	
	\paragraph{Proof of (b)}
	If it holds that $Q_A^{f'\dag}(s,f)=V_A^{f'\dag}(s)~\forall s\in\mathcal{S}$, by the similar derivation of \cref{eq:improve_proof}, we obtain
	\begin{align}
		V_A^{f'\dag}(s_0)=\E^{fR_B^*(f)}\left[\sum_{t=0}^{n-1}\gamma_A^t\hat{r}_A(s_t,a_t,b_t)\middle|s_0\right] + \gamma_A^n\E^{fR_B^*(f)}\left[V_A^{f'\dag}(s_n)\middle|s_0\right]
	\end{align}
	for all $s_0\in\mathcal{S}$.
	Then, in the limit for $n\to\infty$, we have $V_A^{f'\dag}(s_0)=V_A^{f\dag}(s_0)$ for all $s_0\in\mathcal{S}$.
	
	Conversely, if it holds that $V_A^{f'\dag}(s)=V_A^{f\dag}(s)$ for all $s\in\mathcal{S}$, we have
	\begin{subequations}
		\begin{align}
			V_A^{f\dag}(s)&=r_A(s,f(s),R_B^*(s,f))+\gamma_A\E_{s'\sim p(s'|s,f(s),R_B^*(s,f))}\left[V_A^{f\dag}(s')\right]\\
			&=r_A(s,f(s),R_B^*(s,f))+\gamma_A\E_{s'\sim p(s'|s,f(s),R_B^*(s,f))}\left[V_A^{f'\dag}(s')\right]\\
			&=Q_A^{f'\dag}(s,f)
		\end{align}
	\end{subequations}
	for all $s\in\mathcal{S}$, which follows that $Q_A^{f'\dag}(s,f)=V_A^{f\dag}(s)=V_A^{f'\dag}(s)$ for all $s\in\mathcal{S}$.
	
	\paragraph{Proof of (c)} By the definition of $V_A^{f\dag}(s)$, we can rewrite the reward value as
	\begin{equation}
		r_A(s,f(s),R_B^*(s,f)) = V_A^{f\dag}(s) - \gamma_A\E_{s'}\left[V_A^{f\dag}(s')\right].
	\end{equation}
	Plugging it into the definition of $Q_A^{f'\dag}(s,f)$ and subtracting $V_A^{f'\dag}(s)$, we have
	\begin{align}
		Q_A^{f'\dag}(s,f) - V_A^{f'\dag}(s)
		&= r_A(s,f(s),R_B^*(s,f)) + \gamma_A\E_{s'}\left[V_A^{f'\dag}(s')\right] - V_A^{f'}(s) \\
		&= V_A^{f\dag}(s) - V_A^{f'\dag}(s) - \gamma_A\E_{s'}\left[V_A^{f\dag}(s') - V_A^{f'\dag}(s')\right] .
	\end{align}
	The policy improvement is possible if and only if $Q_A^{f'\dag}(\cdot,f) \succ V_A^{f'\dag}$.
	Because of the above equality, equivalently, we can say that the policy improvement from $f'$ is possible if and only if there exists a policy $f$ whose value function satisfies
	\begin{align}
		V_A^{f\dag}(s) - V_A^{f'\dag}(s) \ge \gamma_A\E_{s'}\left[V_A^{f\dag}(s') - V_A^{f'\dag}(s')\right] 
	\end{align}
	for all $s \in \mathcal{S}$, and there exists a state $s$ where the inequality strictly holds.
\end{proof}

\section{Proof of \texorpdfstring{\cref{thrm:lemma_POimplies}}{Lg}}\label{apdx:lemma_POimplies}

\begin{proof}
	We prove the counterpart of \cref{thrm:lemma_POimplies}.
	Taking the negation of \cref{eq:lemma_POimplies} for all $f\in \mathcal{W}_A$, we have
	\begin{subequations}
		\begin{align}
			&\neg\forall f\in\mathcal{W}_A\left\{ Q_A^{f'\dag}(s,f)\ge V_A^{f'\dag}(s)~\forall s\in\mathcal{S}\implies V_A^{f\dag}(s)=V_A^{f'\dag}(s)~\forall s\in\mathcal{S} \right\}\nonumber\\
			&\iff \neg \forall f\in \mathcal{W}_A \left\{ Q_A^{f'\dag}(s,f)<V_A^{f'\dag}(s)~\exists s\in\mathcal{S}\lor V_A^{f\dag}(s)=V_A^{f'\dag}(s)~\forall s\in\mathcal{S} \right\}\\
			&\iff \exists f\in \mathcal{W}_A \left\{ Q_A^{f'\dag}(s,f)\ge V_A^{f'\dag}(s)~\forall s\in\mathcal{S}\land V_A^{f\dag}(s)\neq V_A^{f'\dag}(s)~\exists s\in\mathcal{S} \right\}.
		\end{align}
	\end{subequations}
	From \cref{thrm:lemma_improve} (a), since $V_A^{f\dag}(s)\ge V_A^{f'\dag}(s)~\forall s\in\mathcal{S}$ holds and $Q_A^{f'\dag}(s,f)\ge V_A^{f'\dag}(s)~\forall s\in\mathcal{S}$ holds, we have
	\begin{align}
		&\exists f\in \mathcal{W}_A \left\{ Q_A^{f'\dag}(s,f)\ge V_A^{f'\dag}(s)~\forall s\in\mathcal{S}\land V_A^{f\dag}(s)\neq V_A^{f'\dag}(s)~\exists s\in\mathcal{S} \right\}\nonumber\\
		&\implies \exists f\in \mathcal{W}_A \left\{ V_A^{f\dag}(s)\ge V_A^{f'\dag}(s)~\forall s\in\mathcal{S}\land V_A^{f\dag}(s)\neq V_A^{f'\dag}(s)~\exists s\in\mathcal{S} \right\}
		\\
		&\iff \exists f\in \mathcal{W}_A \left\{ V_A^{f\dag} \succ V_A^{f'\dag} \right\}
		\\
		&\implies \exists v\in \mathcal{V} \subseteq \mathrm{cl}\mathcal{V} \left\{ v \succ V_A^{f'\dag} \right\}
		\\
		&\implies V_A^{f'\dag} \notin \mathcal{PV}.
	\end{align}
	Therefore, $f'$ is not a PO policy. This completes the proof.
\end{proof}

\section{Proof of \texorpdfstring{\cref{thrm:lemma_impliesPO}}{Lg}}\label{apdx:lemma_impliesPO}

\begin{proof}
	First, we prove that (i) $\implies V_A^{f'\dag}(s)>V_A^{f\dag}(s)~\exists s\in\mathcal{S}$.
	Given $f\in\mathcal{W}_A, f'\in\mathcal{W}_A$, by the definition of $\delta(f,f')$, we have
	\begin{equation}
		\E_{s'\sim p(s'|s,f(s),R_B^*(s,f))}\left[ Q_A^{f'\dag}(s',f)-V_A^{f'\dag}(s') \right] \le \delta(f,f')
	\end{equation}
	for all $s\in\mathcal{S}$.
	Thus, we have
	\begin{subequations}
		\begin{align}
			Q_A^{f'\dag}(s,f)&=r_A(s,f(s),R_B^*(s,f))+\gamma_A\E_{s'\sim p(s'|s,f(s),R_B^*(s,f))}\left[ V_A^{f'\dag}(s') \right]\\
			&\ge\begin{multlined}[t]
			    r_A(s,f(s),R_B^*(s,f))\\+\gamma_A\Bigl(\E_{s'\sim p(s'|s,f(s),R_B^*(s,f))}\left[ Q_A^{f'\dag}(s',f) \right]-\delta(f,f')\Bigl)
			\end{multlined}\\
			&\ge\dots\nonumber\\
			&\ge\begin{multlined}[t]
                    \E^{fR_B^*(f)}\left[ \sum_{t=0}^{n-1} \gamma_A^t r_A(s_t,f(s_t),R_B^*(s_t,f))+\gamma_A^n V_A^{f'\dag}(s_n)\middle| s_0=s\right]\\
    			- \left(\sum_{t=0}^{n-1}\gamma_A^t -\gamma_A^0\right)\delta(f,f').
			\end{multlined}
		\end{align}
	\end{subequations}
	for all $s\in\mathcal{S}$.
	In the limit for $n\to\infty$, the first term converges to $V_A^{f\dag}(s)$, the second term converges to 0, and the third term converges to $-\frac{\gamma_A}{1-\gamma_A}\delta(f,f')$.
	Therefore, we have
	\begin{align}
		Q_A^{f'\dag}(s,f)\ge V_A^{f\dag}(s)-\dfrac{\gamma_A}{1-\gamma_A}\delta(f,f')
	\end{align}
	for all $s\in\mathcal{S}$.
	Then, if (i) holds, because there exists $s\in\mathcal{S}$ such that $Q_A^{f'\dag}(s,f)<V_A^{f'\dag}(s)-\frac{\gamma_A}{1-\gamma_A}\delta(f,f')$, it holds that
	\begin{align}
		V_A^{f'\dag}(s)-\frac{\gamma_A}{1-\gamma_A}\delta(f,&f')>Q_A^{f'\dag}(s,f)\ge V_A^{f\dag}(s)-\dfrac{\gamma_A}{1-\gamma_A}\delta(f,f')\quad\exists s\in\mathcal{S}.\\
		&\therefore V_A^{f'\dag}(s)>V_A^{f\dag}(s)\quad\exists s\in\mathcal{S}.\label{eq:(i)implies}
	\end{align}
	If (ii) holds, because of \cref{thrm:lemma_improve} (b), it holds that
	\begin{align}
		V_A^{f'\dag}(s)=V_A^{f\dag}(s)\quad\forall s\in\mathcal{S}.\label{eq:(ii)implies}
	\end{align}
Therefore, it holds that
	\begin{multline}
		\forall f\in \mathcal{W}_A \{\mathrm{(i)~or~(ii)}\}\\
		\implies \forall f\in \mathcal{W}_A \left\{ V_A^{f'\dag}(s)>V_A^{f\dag}(s)~~\exists s\in\mathcal{S} \mathrm{~~or~~} V_A^{f'\dag}(s)=V_A^{f\dag}(s)~~\forall s\in\mathcal{S}\right\}\label{eq:def_POP_alt},
	\end{multline}
	where the right-hand side is equivalent to the definition of PO policies (\cref{thrm:def_POP}).
\end{proof}

\section{Proof of \texorpdfstring{\cref{thrm:thrm_algo}}{Lg}}\label{apdx:thrm_algo}

\begin{proof}
	We prove each statement one by one. 
    The proof of (a) is based on the monotonicity of $\{v_t\}$ and the compactness of $\mathrm{cl}\mathcal{V}$.
    The proof of (b) is based on the statement (b) of \Cref{thrm:lemma_improve}.
    The statements (c) and (d) are proved by contradiction using the statement (c) of \Cref{thrm:lemma_improve}.
    
	\paragraph{Proof of (a)}
	As $f_{t+1}\in\mathcal{W}_\succcurlyeq(f_t)$, it holds that
	\begin{align}
		v_{t+1} = V_A^{f_{f+1}\dag} \succcurlyeq
		Q_A^{f_t\dag}(\cdot,f_{t+1}) \succcurlyeq
		V_A^{f_t\dag}(\cdot) = v_t. \label{eq:monotone_improve_2}
	\end{align}
	Thus, $v_{t+1} \succcurlyeq v_{t}$ holds for all $t\in\N$.
	
	Because $r_A$ is bounded, so is $V_A^{f\dag}$, implying $\mathrm{cl}\mathcal{V}$ is compact. 
	Considering the sequence $\{v_t(s)\}_{t=0}^{\infty}$ for each $s \in \mathcal{S}$, it is a monotonically increasing sequence. $\{v_t(s)\}_{t=0}^{\infty}$ converges in $\mathrm{cl}\mathcal{V}$ because $\mathrm{cl}\mathcal{V}$ is compact. 
	Let $v_{\infty}(s) = \lim_{t\to\infty} v_t(s)$ for each $s \in \mathcal{S}$. 
	Then, it holds that $v_\infty = \lim_{t\to\infty} v_t$. 
	
	\paragraph{Proof of (b)}
	First, $v_{t+1} \eqs v_t \impliedby v_t \eqs v_\infty$ is apparent.
	
	Second, we prove $v_{t+1} \eqs v_{t} \implies v_t \eqs v_\infty$.
	As $f_{t+1}\in\mathcal{W}_{1-\epsilon}(f_t)$, it holds that
	\begin{align}
		\mathcal{L} \left[ V_A^{f_{t+1}\dag} \right] - \mathcal{L}\left[ V_A^{f_t\dag} \right]
		&\ge 
		\mathcal{L} \left[ Q_A^{f_t\dag}(\cdot,f_{t+1}) \right] - \mathcal{L}\left[ V_A^{f_t\dag} \right]\\
		&\ge (1 - \epsilon) \sup_{f \in \mathcal{W}_\succcurlyeq(f_t)} \Bigl(\mathcal{L}\left[ Q_A^{f_t\dag}(\cdot,f) \right] - \mathcal{L}\left[ V_A^{f_t\dag} \right]\Bigl). \label{eq:monotone_improve_3}
	\end{align}
	If $v_{t+1} \eqs v_{t}$, we have $\mathcal{L}[v_{t+1}]-\mathcal{L}[v_{t}] = 0$; hence, the above inequality implies
	\begin{align}
		\sup_{f \in \mathcal{W}_\succcurlyeq(f_t)} \mathcal{L}\left[ Q_A^{f_t\dag}(\cdot,f) \right] = \mathcal{L}\left[ V_A^{f_t\dag} \right]; 
	\end{align}
	therefore, for all $f \in \mathcal{W}_\succcurlyeq(f_t)$, we have 
	$\mathcal{L}\left[ Q_A^{f_t\dag}(\cdot,f) \right] = \mathcal{L}\left[ V_A^{f_t\dag} \right]$.
	Because $\mathcal{L}$ is Pareto-compliant, the above equality implies 
	$Q_A^{f_t\dag}(\cdot,f) \not \succ V_A^{f_t\dag}$.
	However, because $f \in \mathcal{W}_\succcurlyeq(f_t)$, we have $Q_A^{f_t\dag}(\cdot,f) \succcurlyeq V_A^{f_t\dag}$, implying $Q_A^{f_t\dag}(\cdot,f) \eqs V_A^{f_t\dag}$.
	From Statement (b) of \cref{thrm:lemma_improve}, we have $V_A^{f\dag} \eqs V_A^{f_t\dag}$ for all $f \in \mathcal{W}_\succcurlyeq(f_t)$.
	(Conversely, if $V_A^{f\dag} \eqs V_A^{f_t\dag}$ for all $f \in \mathcal{W}_\succcurlyeq(f_t)$, we have $v_{t+1} \eqs v_{t}$.)
	Thus, $\mathcal{W}_\succcurlyeq(f)=\mathcal{W}_\succcurlyeq(f_t)$ holds for all $f\in\mathcal{W}_\succcurlyeq(f_t)$.
	It implies that $\mathcal{W}_\succcurlyeq(f_{t+1})=\mathcal{W}_\succcurlyeq(f_t)$ since $f_{t+1}\in\mathcal{W}_\succcurlyeq(f_t)$; thus, $f_{t+2}\in\mathcal{W}_\succcurlyeq(f_{t+1})=\mathcal{W}_\succcurlyeq(f_t)$.
	Therefore, 
	\begin{align}
		V_A^{f_t\dag} \eqs V_A^{f_{t+1}\dag} \eqs V_A^{f_{t+2}\dag} \eqs \cdots
	\end{align}
	and hence $v_{t+k} \eqs v_{t}$ for all $k \ge 0$, implying that $v_{t} = v_\infty$.    
	
	\paragraph{Proof of (c)}
	First, we show that $$\min_{s \in \mathcal{S}} (v(s) - v_\infty(s))  \le \gamma_A( \max_{s \in \mathcal{S}} (v(s) - v_\infty(s)) )$$ for all $v\in \mathcal{V}$.
	
	Suppose that there exists $v^* \in \mathcal{V}$ such that $$\min_{s \in \mathcal{S}} (v^*(s) - v_\infty(s)) > \gamma_A(\max_{s \in \mathcal{S}} (v^*(s) - v_\infty(s)))$$ holds. 
	Then, $v^* \succ v_\infty$ and there exists $f^* \in \mathcal{W}_{A}$ whose value function is $V_A^{f^*\dag} \eqs v^*$. Let $\tau = \frac{ \min_{s \in \mathcal{S}} (v^*(s) - v_\infty(s)) }{ \max_{s \in \mathcal{S}} (v^*(s) - v_\infty(s)) }$. We have $1 \ge \tau > \gamma_A$. 
	
	For an arbitrarily small $\xi > 0$ there exists $t$ such that $\norm{v_t - v_\infty}_{\infty} \le \xi$ because $v_\infty \eqs \lim_{t\to\infty}v_t$.
	Because of the continuity of $\mathcal{L}$, it also implies that for an arbitrarily small $\xi' > 0$ there exists $t'$ such that $\mathcal{L}[v_{t'}] > \mathcal{L}[v_\infty] - \xi'$. 
	Therefore, we can find $t$ satisfying
	\begin{align}
		\norm{v_\infty - v_t}_\infty &\le \delta \frac{\tau - \gamma_A}{\gamma_A} \norm{v^{*} - v_\infty}_{\infty} \quad \text{for some $\delta \in (0, 1)$}, \quad \text{and} \label{eq:t-cond-2}
		\\
		\mathcal{L}[v_\infty] - \mathcal{L}[v_t] &< (1 - \epsilon) (\mathcal{L}[\bar{v}] - \mathcal{L}[v_\infty]), \label{eq:t-cond-1}
	\end{align}
    where $\bar{v}(s) := v_\infty(s) + (1 - \delta)(\tau - \gamma_A) \norm{v^{*} - v_\infty}_{\infty}.$
	Let $f_t$ be the corresponding policy whose value function is $V_A^{f_t\dag} \eqs v_t$. Here, we assume that $\gamma_A > 0$. For the case of $\gamma_A = 0$, Statement (c) is an immediate consequence of Statement (d) because $v_{\infty}\in\mathcal{PV}$ and $\mathcal{PV} \subseteq \partial \mathcal{V}$. 
	
	Subsequently, we show $V_A^{f^*\dag} \succcurlyeq Q_A^{f_t\dag}(\cdot, f^*) \succcurlyeq \bar{v}_\infty$ using \cref{eq:t-cond-2}. 
	In the proof of Statement (c) of \cref{thrm:lemma_improve}, for all $s \in \mathcal{S}$,
	\begin{equation}
		Q_A^{f'\dag}(s,f) - V_A^{f'\dag}(s)
		= V_A^{f\dag}(s) - V_A^{f'\dag}(s) - \gamma_A\E_{s'}\left[V_A^{f\dag}(s') - V_A^{f'\dag}(s')\right] .
	\end{equation}
	Notably, $\E_{s'}\left[V_A^{f\dag}(s') - V_A^{f'\dag}(s')\right] \le \norm{V_A^{f\dag} - V_A^{f'\dag}}_{\infty}$. 
	Letting $f = f^*$ and $f' = f_t$, we have, for all $s \in \mathcal{S}$,
	\begin{align}
		Q_A^{f_t\dag}(s,f^*) - V_A^{f_t\dag}(s)
		&\ge V_A^{f^*\dag}(s) - V_A^{f_t\dag}(s) - \gamma_A \norm{V_A^{f^*\dag} - V_A^{f_t\dag}}_{\infty}.
	\end{align}
	By adding $V_A^{f_t\dag}(s) - v_{\infty}(s)$ to both sides of the above inequality, we obtain, for all $s \in \mathcal{S}$,
	\begin{subequations}
		\begin{align}
			Q_A^{f_t\dag}(s,f^*) - v_{\infty}(s)
			&\ge v^*(s) - v_{\infty}(s) - \gamma_A \norm{v^* - v_t}_{\infty} \\
			&\ge \min_{s \in \mathcal{S}} \left( v^*(s) - v_{\infty}(s)\right) - \gamma_A \norm{v^* - v_t}_{\infty} \\
			&= \min_{s \in \mathcal{S}} \left( v^*(s) - v_{\infty}(s)\right) - \gamma_A \norm{v^* - v_{\infty} + v_{\infty} - v_t}_{\infty} \\
			&\ge \min_{s \in \mathcal{S}} \left( v^*(s) - v_{\infty}(s)\right) - \gamma_A \norm{v^* - v_{\infty}}_\infty - \gamma_A \norm{v_{\infty} - v_t}_{\infty} \\
			&= (\tau - \gamma_A) \norm{v^* - v_{\infty}}_{\infty} - \gamma_A \norm{v_{\infty} - v_t}_{\infty} \\
			&\ge (1 - \delta)(\tau - \gamma_A) \norm{v^* - v_{\infty}}_{\infty}.
		\end{align}
	\end{subequations}
	Therefore, we have $Q_A^{f_t\dag}(\cdot, f^*) \succcurlyeq \bar{v}$.
	Because $\bar{v} \succ v_{\infty} \succcurlyeq V_A^{f_t\dag}$, it implies that $Q_A^{f_t\dag}(\cdot, f^*) \succcurlyeq V_A^{f_t\dag}$. From Statement (a) of \cref{thrm:lemma_improve}, we have $V_A^{f^*\dag} \succcurlyeq Q_A^{f_t\dag}(\cdot, f^*)$. 
	
	Finally, we derive a contradiction.
	Let $\ell_{t}^{\sup} = \sup_{f \in \mathcal{W}_\succcurlyeq(f_t)} \mathcal{L}[Q_A^{f_t\dag}(\cdot, f)]$. 
	Because $f^* \in \mathcal{W}_\succcurlyeq(f_t)$, we have $\ell_{t}^{\sup} \ge \mathcal{L}[Q_A^{f_t\dag}(\cdot, f^*)]$.
	We also know that $\mathcal{L}[Q_A^{f_t\dag}(\cdot, f^*)] \ge \mathcal{L}[\bar{v}]$. 
	Because $f_{t+1} \in \mathcal{W}_{1-\epsilon}(f_t)$, we obtain
	\begin{subequations}
		\begin{align}
			\mathcal{L}[v_{t+1}] - \mathcal{L}[v_{t}]
			&\ge \mathcal{L}[Q_A^{f_t\dag}(\cdot, f_{t+1})] - \mathcal{L}[v_{t}] \\
			&\ge (1 - \epsilon) \left(\sup_{f \in \mathcal{W}_\succcurlyeq(f_t)} \mathcal{L}[Q_A^{f_t\dag}(\cdot, f)] - \mathcal{L}[v_{t}]\right) \\
			&\ge (1 - \epsilon) ( \mathcal{L}[Q_A^{f_t\dag}(\cdot, f^*)] - \mathcal{L}[v_\infty]) \\            
			&\ge (1 - \epsilon) (\mathcal{L}[\bar{v}] - \mathcal{L}[v_\infty]) \\            
			&> \mathcal{L}[v_\infty] - \mathcal{L}[v_t] ,
		\end{align}
	\end{subequations}
	where we used \cref{eq:t-cond-1} for the last inequality.
	It implies $\mathcal{L}[v_{t+1}] > \mathcal{L}[v_\infty]$, which contradicts to the fact that $v_\infty \succcurlyeq v_{t+1}$.
	Therefore, we have $\min_{s \in \mathcal{S}} (v(s) - v_\infty(s))  \le \gamma_A\left( \max_{s \in \mathcal{S}} (v(s) - v_\infty(s)) \right)$ for all $v\in \mathcal{V}$. 
	
	Subsequently, we show that $v_\infty \in \partial \mathcal{V}$. 
	
	Suppose that $v_\infty \in \mathcal{V}\setminus \partial\mathcal{V}$ (i.e., $v_\infty$ is an interior of $\mathcal{V}$). 
	Then, there exists an $r > 0$ such that $\mathcal{E} = \{v \in \mathcal{F}_\mathcal{S} \mid \norm{v - v_\infty}_{\infty} \le r\} \subseteq \mathcal{V}$.
	Let $v_{\infty}^r$ be such that $v_{\infty}^r(s) = v_{\infty}(s) + r$ for all $s\in\mathcal{S}$. 
	Then, $v_{\infty}^r \in \mathcal{E}$. 
	Moreover, because $v_{\infty}^r \in \mathcal{V}$, there exists a policy $f_\infty^r$ such that $V_A^{f_\infty^r \dag} = v_{\infty}^r$.
	
	We lead to the contradiction to $\min_{s \in \mathcal{S}} v(s) - v_\infty(s)  \le \gamma_A\left( \max_{s \in \mathcal{S}} v(s) - v_\infty(s) \right)$.
	Using \cref{eq:t-cond-2}, we obtain
	\begin{equation}
		\frac{\min_{s \in \mathcal{S}} V_A^{f_\infty^r \dag}(s) - v_\infty(s)}{\max_{s \in \mathcal{S}} V_A^{f_\infty^r \dag}(s) - v_\infty(s)} 
		= 
		\frac{\min_{s \in \mathcal{S}} (v_\infty^r(s) - v_\infty(s)) }{\max_{s \in \mathcal{S}} (v_\infty^r(s) - v_\infty(s))} 
		= 1,
	\end{equation}
	contradicting $\min_{s \in \mathcal{S}} v(s) - v_\infty(s)  \le \gamma_A\left( \max_{s \in \mathcal{S}} v(s) - v_\infty(s) \right)$ for all $v \in \mathcal{V}$. 
	
	\paragraph{Proof of (d)}
	In case of $\gamma_A = 0$, we have $Q_A^{f_t\dag}(\cdot,f) \eqs V_A^{f\dag}$ for all $f_t, f\in\mathcal{W}_A$.

	Suppose that $v_\infty \notin \mathcal{PV}$. 
	Then, there exists a value function $v^*\in\mathrm{cl}\mathcal{V}$ such that $v^* \succ v_\infty$.
	Similar to the proof of Statement (c), we choose $t$ such that
	\begin{equation}
		\mathcal{L}[v_\infty] - \mathcal{L}[v_t] < (1 - \epsilon) (\mathcal{L}[v^*] - \mathcal{L}[v_\infty]).
	\end{equation}
	Let $\mathcal{V}_\succcurlyeq(v_t):=\{v\in\mathcal{V}\mid v\succcurlyeq v_t\}$ and $\mathrm{cl}\mathcal{V}_\succcurlyeq(v_t):=\{v\in\mathrm{cl}\mathcal{V}\mid v\succcurlyeq v_t\}$.
	Then, we have $v^* \in \mathrm{cl}\mathcal{V}_\succcurlyeq(v_t)$ because $v^* \succ v_\infty \succcurlyeq v_t$.
	Moreover, because $\mathcal{V}_\succcurlyeq(v_t) = \{V_A^{f\dag} : f \in \mathcal{W}_{\succcurlyeq}(f_t)\}$, 
	we obtain
	\begin{subequations}
		\begin{align}
			\ell_t^{\sup} :=& \sup_{f \in \mathcal{W}_\succcurlyeq(f_t)} \mathcal{L}[V_A^{f\dag}] \\
			=& ~\sup_{v\in\mathcal{V}_\succcurlyeq(v_t)} \mathcal{L}[v]\\
			=& \max_{v\in\mathrm{cl}\mathcal{V}_\succcurlyeq(v_t)} \mathcal{L}[v]\\
			\ge&~ \mathcal{L}[v^*].
		\end{align}
	\end{subequations}
	Therefore, 
	\begin{subequations}
		\begin{align}
			\mathcal{L}[v_{t+1}] - \mathcal{L}[v_{t}]
			&\ge (1 - \epsilon) (\ell_t^{\sup} - \mathcal{L}[v_{t}]) \label{eq:68a}\\
			&\ge (1 - \epsilon) (\ell_t^{\sup} - \mathcal{L}[v_\infty]) \\            
			&\ge (1 - \epsilon) (\mathcal{L}[v^*] - \mathcal{L}[v_\infty]) \\            
			&> \mathcal{L}[v_\infty] - \mathcal{L}[v_t] ,
		\end{align}
	\end{subequations}
	where the derivation of \cref{eq:68a} is due to the definition of $\mathcal{W}_{1-\epsilon}(f_t)$ in \cref{eq:Wcond} and $Q_A^{f_t\dag}(\cdot,f) \eqs V_A^{f\dag}$. 
	This implies $\mathcal{L}[v_{t+1}] > \mathcal{L}[v_\infty]$, which contradicts to $v_\infty \succcurlyeq v_{t+1}$. 
	Therefore, $v_\infty \in \mathcal{PV}$.
\end{proof}

\section{Splitting the Search Space}\label{sec:improve_algo_feasibility}

The proposed algorithm in \cref{algo:proposed} can be impractical when the leader's policy space $\mathcal{W}_A$ is prohibitively large to search for the next policy in \cref{eq:def_hat_W} and \cref{eq:update}.
To avoid the exhaustive search, we propose a splitting strategy for the search space to obtain improved policies from each subspace efficiently.

We first split $\mathcal{W}_A$ into disjoint sets $\mathcal{W}_{1} \cup \dots \cup \mathcal{W}_{K}$, where the best response is $R_B^*(f) = g_i^* \in \mathcal{W}_B^d$ for $f \in \mathcal{W}_{i}$. Because the set of deterministic policies, $\mathcal{W}_B^d$, is finite, we can always find such a separation. Additionally, we argue that the number of deterministic policies that form the best response is rather limited, enabling us to enumerate them easily.

We consider replacing \eqref{eq:update} with locating a Pareto optimal $f$ with respect to $Q_A^{f_t\dagger}(\cdot, f)$. Because \eqref{eq:update} also locates Pareto optimal $f$, this replacement corresponds to selecting a different $\mathcal{L}$ at each $t$.

This is realized in two steps. First, we locate a policy satisfying $Q_A^{f_t\dagger}(\cdot, f) \succcurlyeq V_A^{f_t\dagger}$. It is nontrivial as the best response $R_B^*$ changes as $f$ changes. However, if we limit our attention to a subset $\mathcal{W}_k$ for some $k$, the best response does not change with $f \in \mathcal{W}_k$. Then, locating $f \in \mathcal{W}_k$ satisfying $Q_A^{f_t\dagger}(\cdot, f) \succcurlyeq V_A^{f_t\dagger}$ is casted as the following constrained optimization problem:
\begin{align}
\min_{z, f \in \mathcal{W}_k} z \quad \text{s.t.}\quad V_A^{f_t\dag}(s) - Q_A^{f_t\dagger}(s, f) \leq z \quad \text{for all } s \in \mathcal{S},\label{eq:practical_derivation1}
\end{align}
where $Q_A^{f_t\dagger}(s, f)$ is differentiable with respect to $f$ and is easily computable as the best response is constant. Therefore, it will be solved by, e.g., a gradient-based solver with projection onto $\mathcal{W}_k$. 
A solution with $z \le 0$ satisfies $Q_A^{f_t\dagger}(\cdot, f) \succcurlyeq V_A^{f_t\dagger}$. Such a solution must exist in $\mathcal{W}_k$ for some $k$ because $f_t$ satisfies this condition. 

Once we find such a solution, denoted as $(z_t, f_t')$, we find a Pareto optimal policy for $Q_A^{f_t\dagger}(\cdot, f)$ in $\mathcal{W}_k \cap \mathcal{W}_{\succcurlyeq}(f_t)$ by improving $f_t'$. It is done by performing a Pareto ascent method with projection, 
\begin{equation}
f \gets \Pi_{\mathcal{W}_k}( f + \eta \Delta ), \quad \text{where} \quad \Delta^{\top} \nabla Q_A^{f_t\dagger}(s, f) \geq 0 \quad \text{for all } s \in \mathcal{S},\label{eq:pareto-ascent}
\end{equation}
or its variants, e.g., \citet{Harada2006pdm}, with $f_t$ as a feasible initial solution of this step.
If we find $f_t^*$ such that $Q_A^{f_t\dagger}(\cdot, f) \succ V_A^{f_t\dagger}$, we let $f_{t+1} = f_t^*$. These steps will be continued until a predefined termination criterion is satisfied, or no policy strictly dominating the current policy is found. 
The whole algorithm is shown in \cref{algo:proposed-practical}.

\begin{algorithm}[H]
	\caption{Practical Pareto-Optimal Policy Iteration}
	\label{algo:proposed-practical}
	\begin{algorithmic}[1]
		\REQUIRE{Maximum number of iterations $M$.}
        \STATE {Split the leader's policy space $\mathcal{W}_A$ into disjoint sets $\mathcal{W}_{1} \cup \dots \cup \mathcal{W}_{K}$, where the best response is $R_B^*(f) = g_i^* \in \mathcal{W}_B^d$ for $f \in \mathcal{W}_{i}$}
		\STATE {Randomly initialize $f_0\in\mathcal{W}_A$ and compute $g_0 = R_B^*(f_0)$}
		\FOR {$t=0$ to $M-1$}
		\STATE {Compute $V_A^{f_t\dag}$ by repeatedly applying $T_A^{f_tR_B^*(f_t)}$}
        \STATE {Let $\mathcal{I}_t = \{1, \dots, K\}$}
        \WHILE {$\abs{\mathcal{I}_t} > 0$}
        \STATE {Randomly select $k \in \mathcal{I}_t$}
        \STATE {Solve \eqref{eq:practical_derivation1} to obtain $z_t, f_t'$}
        \IF {  $z_t \le 0$ }
        \STATE {Perform the Pareto ascent \eqref{eq:pareto-ascent} to obtain $f_{t}^*$}
        \STATE {Let $f_{t+1} = f_t^*$ and {\bf break if} $Q_A^{f_t\dagger}(\cdot, f_t^*) \succ V_A^{f_t\dag}$}
        \ENDIF
        \STATE Update $\mathcal{I}_t \gets \mathcal{I}_t\setminus\{k\}$
        \ENDWHILE
        \STATE { Let $f_{M} = \dots = f_{t+1} = f_{t}$ and {\bf break if} $\mathcal{I}_t = \emptyset$}
		\ENDFOR
		\ENSURE{A stationary policy $f_M$}
	\end{algorithmic}
\end{algorithm}

\section{Backtracking}\label{apdx:proposed_2}

\Cref{algo:proposed} may not output a PO policy because the terminate condition in line \ref{algo:proposed:1_l6} of \cref{algo:proposed} is the necessary condition for PO policies.
In this case, \cref{thrm:lemma_impliesPO} can be used to determine whether the output policy is a PO policy: provided (i) or (ii) in \cref{thrm:lemma_impliesPO} holds for all $f\in\mathcal{W}_A$ with the output policy $f_t$ and its state value function $V_A^{f_t\dag}$, $f_t$ is a PO policy.
Notably, this judgment is not perfect because the policy may be determined not to be a PO policy when it really is a PO policy. 

We proposed selecting another policy $f'\in\mathcal{W}(f_{t-1})$ that has not been selected thus far and restarting the for loop with $f_t\leftarrow f'$, which we call \textit{"Backtracking"}, to deal with the case where the output policy $f_t(t<M)$ is determined not to be a PO policy.
If there is no policy left in $\mathcal{W}(f_{t-1})$, then it backtracks one by one, as $\mathcal{W}(f_{t-2}), \mathcal{W}(f_{t-3}), \dots$, until it reaches the set with policies that have never been selected.
This method requires that $\mathcal{W}(f_t)$ is saved at each iteration.

The entire proposed algorithm with the PO-policy judgement and the backtracking is shown in \cref{algo:proposed_2}.
$W$ is a list to save each $\mathcal{W}(f_t)$ and $M$ is the maximum number of iterations.
This algorithm is designed such that the number of times to compute $V_A^{f_t\dag}$ in line \ref{algo:l5} and $\mathcal{W}_\succcurlyeq(f_t)$ in line \ref{algo:l7} is at most $M$.
If the PO-policy judgment in line \ref{algo:l9} does not return \texttt{True} during $M$ iterations, the algorithm outputs the policy that maximizes the objective $\sum_{s\in\mathcal{S}}\alpha_sV_A^{f_t\dag}(s)$ among the ones that satisfy the necessary condition for PO policies thus far and $f_M$ (see lines \ref{algo:l12}, \ref{algo:l25}, and \ref{algo:l26}).

\begin{algorithm}[H]
	\caption{Pareto-Optimal Policy Iteration with Backtracking}
	\label{algo:proposed_2}
	\begin{algorithmic}[1]
		\REQUIRE{Pareto-compliant scalarization $\mathcal{L}$, maximum number of iterations $M$, a $M$-length list of empty sets $W$.}
		\STATE {$W(0)\leftarrow\mathcal{W}_A$}
		\STATE {Randomly sample $f_0$ from $W(0)$ without replacement.}
		\STATE {$\mathcal{W}_{output}\leftarrow \{\}$}
		\FOR {$t=0$ to $M-1$}
		\STATE {Compute $V_A^{f_t\dag}\in\mathcal{F}(\mathcal{S})$ such that $(T_A^{f_tR_B^*(f_t)}V_A^{f_t\dag})(s)=V_A^{f_t\dag}(s)$ for all $s\in\mathcal{S}$.}\alglinelabel{algo:l5}
		\STATE {$Q_A^{f_t\dag}(s,a,b)\leftarrow r_A(s,a,b)+\gamma_A\E_{s'\sim p(s'|s,a,b)}\left[V_A^{f_t\dag}(s')\right]$ for all $s\in\mathcal{S}, a\in\mathcal{A}, b\in\mathcal{B}$.}
		\STATE {$\mathcal{W}_\succcurlyeq(f_t)\leftarrow\{f\in\mathcal{W}_A~|~\E_{a\sim f(s)}\left[Q_A^{f_t\dag}(s,a,R_B^*(s,f))\right]\ge V_A^{f_t\dag}(s)~\forall s\in\mathcal{S}\}$.}\alglinelabel{algo:l7}
		\IF{$\E_{a\sim f(s)}\left[Q_A^{f_t\dag}(s,a,R_B^*(s,f))\right]=V_A^{f_t\dag}(s)~\forall s\in\mathcal{S}, \forall f\in\mathcal{W}_\succcurlyeq(f_t)$}
		\IF{$f_t$ satisfies (i) or (ii) of \cref{thrm:lemma_impliesPO} for all $f\in\mathcal{W}_A$}\alglinelabel{algo:l9}
		\STATE {{\bf return} $f^*\leftarrow f_t$.}
		\ENDIF
		\STATE {Put $f_t$ into $\mathcal{W}_{output}$.}\alglinelabel{algo:l12}
		\STATE \COMMENT{Backtracking}
		\STATE {$k\leftarrow t$}
		\WHILE{$|W(k)|=0$}
		\STATE {$k\leftarrow k-1$}
		\ENDWHILE
		\STATE {Randomly sample $f_{t+1}$ from $W(k)$ without replacement.}
		\ELSE
		\STATE {$\mathcal{W}(f_t)\leftarrow\argmax_{f\in\mathcal{W}_\succcurlyeq(f_t)} \mathcal{L}\left[ \E_{a\sim f(\cdot)}\left[Q_A^{f_t\dag}(\cdot,a,R_B^*(\cdot,f))\right] \right]$.}
		\STATE {$W(t+1)\leftarrow \mathcal{W}(f_t)$.}
		\STATE {Randomly sample $f_{t+1}$ from $W(t+1)$ without replacement.}
		\ENDIF
		\ENDFOR
		\STATE {Put $f_M$ into $\mathcal{W}_{output}$.}\alglinelabel{algo:l25}
		\STATE {{\bf return} $f^*\in\argmax_{f_t\in\mathcal{W}_{output}}\mathcal{L} \left[ V_A^{f_t\dag}(\cdot)\right]$}
		\alglinelabel{algo:l26}
		\ENSURE{A stationary policy $f^*$}
	\end{algorithmic}
\end{algorithm}

\section{Relation between the methods of Zhang et al. (2020) and Bucarey et al. (2022)}\label{apdx:Zhang}

\citet{Zhang2020-hw} defined the optimal value functions as $V_i^* (i\in\{A,B\})$, which are the solution of the following equations \citep{Zhang2020-hw}: let $a\in\mathcal{A}, b\in\mathcal{B}, \gamma\in[0,1)$,  and for all $s\in\mathcal{S}$,
\begin{align}
	&Q_i^*(s,a,b) = r_i(s,a,b) + \gamma \E_{s'\sim p(s'|s,a,b)} \left[V_i^*(s')\right],\label{eq:zhang_fp_1}\\
	&V_i^*(s) = \textsc{Stackelberg}_i(Q_A^*(s),Q_B^*(s)),\label{eq:zhang_fp}
\end{align}
where $Q_i^*(s)$ is the Q-value $Q_i(a,b,s)$ of an action pair $(a,b)\in\mathcal{A}\x\mathcal{B}$ in state $s$.
Given the fixed $s$, $Q_i^*(b,a,s)$ can be viewed as the payoff of $(a,b)$ for the agent $i$; thus, $Q_i^*(s)$ can be viewed as the payoff table of agent $i$ in the state $s$.
$\textsc{Stackelberg}_i(Q_A^*(s),Q_B^*(s))$ is the payoff for agent $i$ in the Stackelberg equilibrium of the normal-form game represented by the payoff tables $(Q_A^*(s),Q_B^*(s))$.
Therefore, \cref{eq:zhang_fp} can be written as
\begin{subequations}\label{eq:zhang_fp_0}
	\begin{align}
		V_i^*(s) &= Q_i^*(s,R_A'(s),R_B'(s,R_A'(s))),\\
		\where &R_A'(s) := \argmax_{a\in\mathcal{A}}Q_A^*(s,a,R_B(s,a)),\\
		&R_B'(s,a) := \argmax_{b\in\mathcal{B}}Q_B^*(s,a,b).
	\end{align}
\end{subequations}
By substituting \cref{eq:zhang_fp_1} to \cref{eq:zhang_fp_0}, we obtain
\begin{subequations}\label{eq:zhang_fp_2}
	\begin{align}
		V_i^*(s) &= r_i(s,R_A'(s),R_B'(s,R_A'(s))) + \gamma \E_{s'\sim p(s'|s,R_A'(s),R_B(s,R_A'(s)))} \left[V_i^*(s')\right],\\
		\where &R_A'(s) := \argmax_{a\in\mathcal{A}}r_A(s,a,R_B'(s,a)) + \gamma \E_{s'\sim p(s'|s,a,R_B(s,a))} \left[V_A^*(s')\right],\\
		&R_B'(s,a) := \argmax_{b\in\mathcal{B}}r_B(s,a,b) + \gamma \E_{s'\sim p(s'|s,a,b))} \left[V_B^*(s')\right].
	\end{align}
\end{subequations}
Meanwhile, the operator used in \citet{Bucarey2022-ro} is given by
\begin{subequations}\label{eq:quasi-SEeq_Bucarey_apdx}
	\begin{align}
		(Tv)_i(s)&=\begin{multlined}[t]
		    r_i(s,R_A(s,v),R_B(s,R_A(s,v),v_B)) \\
                + \gamma\E_{s'\sim p(s'|s,R_A(s,v),R_B(s,R_A(s,v),v_b))}\left[v_i(s')\right],
		\end{multlined}\\
		\where 
		&R_A(s,v):=\begin{multlined}[t]
		    \argmax_{f_s\in\Delta(\mathcal{A})} r_A(s,f_s,R_B(s,f_s,v_B)) \\
                + \gamma\E_{s'\sim p(s'|s,f_s,R_B(s,f_s,v_B))}\left[v_A(s')\right],
		\end{multlined}\\
		&R_B(s,f_s,v_B):=\argmax_{b\in\mathcal{B}} r_B(s,f_s,b) + \gamma\E_{s'\sim p(s'|s,f_s,b)}\left[v_B(s')\right].
	\end{align}
\end{subequations}
Notably, $\gamma=\gamma_A=\gamma_B$.
Let us consider only the deterministic leader policies in the fixed point of \cref{eq:quasi-SEeq_Bucarey_apdx} and replace $\argmax_{f_s\in\Delta(\mathcal{A})}$ with $\argmax_{a\in\mathcal{A}}$ in the definition of $R_A$.
Then, letting $(V_A,V_B)$ be the fixed point of $T$, it holds from \cref{eq:quasi-SEeq_Bucarey_apdx}, for all $s\in\mathcal{S}$, which is expressed as
\begin{subequations}\label{eq:quasi-SEeq_Bucarey_apdx_2}
	\begin{align}
		V_i(s)&=\begin{multlined}[t]
		    r_i(s,R_A(s,V),R_B(s,R_A(s,V),V_B)) \\
                + \gamma\E_{s'\sim p(s'|s,R_A(s,V),R_B(s,R_A(s,V),V_B))}\left[V_i(s')\right],
		\end{multlined}\\
		\where 
		&R_A(s,V):=\begin{multlined}[t]
		    \argmax_{a\in\mathcal{A}} r_A(s,a,R_B(s,a,V_B)) \\
                + \gamma\E_{s'\sim p(s'|s,a,R_B(s,a,V_B))}\left[V_A(s')\right],
		\end{multlined}\\
		&R_B(s,a,V_B):=\begin{multlined}[t]
		    \argmax_{b\in\mathcal{B}} r_B(s,a,b) 
                + \gamma\E_{s'\sim p(s'|s,a,b)}\left[V_B(s')\right].
		\end{multlined}
	\end{align}
\end{subequations}
Comparing these with \cref{eq:zhang_fp_2}, the definition of $(V_A^*,V_B^*)$ by \cref{eq:zhang_fp_2} is equivalent to the definition of $(V_A,V_B)$ by \cref{eq:quasi-SEeq_Bucarey_apdx_2}.
It follows that the fixed point in Zhang et al. (2020) is the same in Bucarey et al. (2022) when the leader policies in the fixed point are restricted to deterministic stationary policies.

By applying the discussion in the proof of \cref{thrm:thrm_Bucarey} to $(V_A,V_B)$ defined by \cref{eq:quasi-SEeq_Bucarey_apdx_2}, we obtain
\begin{align}
	V_A(s)=\max_{f\in\mathcal{W}_A^d}V_A^{fR_B(f,V_B)}(s),\quad V_B(s)=\max_{g\in\mathcal{W}_B}V_B^{R_A(V)g}(s),
\end{align}
for all $s\in\mathcal{S}$.
Then, the problem of the method of Bucarey et al. discussed in \cref{sec:BucareyMethod} also occurs to $(V_A,V_B)$: $V_A$ does not coincide with the state value function of the deterministic SSE policies since $R_B(f,V_B)$ is not guaranteed to be the best response against any $f\in\mathcal{W}_A^d$.

\section{Application: Policy Teaching by Intervention to the Transitions}\label{sec:application}

In single-agent MDPs, the policy aimed by the agent can be changed by making changes to the agent's observations, reward signals, and transition probabilities, among others.
For example, in an MDP $(\mathcal{S},\mathcal{A},p,\rho,r,\gamma)$, the changes in the observations and the transition probabilities are represented by replacement to another transition function $p'$, which changes the agent's aim from the optimal policies of $(\mathcal{S},\mathcal{A},p,\rho,r,\gamma)$ to those of $(\mathcal{S},\mathcal{A},p',\rho,r,\gamma)$.
This is the theoretical foundation of poisoning attacks against RL~\citep{Behzadan2017-vs,Huang2019-bh,Zhang2020-ai,Rakhsha2020-sc,Sun2020-bo} and policy teaching~\citep{Zhang2008-ll,Zhang2009-oz}, which aims to guide the agent's policy learning.

SSGs can be viewed as a model of intervention in transition probabilities by the leader, where the follower is the victim of the attack or the teaching. 
Even if the follower takes the same action at each state, the transition can be shifted by the leader changing its actions because the transition probability depends on the actions of all the agents.
Therefore, suppose the follower cannot observe both the attendance and the actions of the leader, then the follower's learning can be represented as learning the optimal policies of the single-agent MDP $(\mathcal{S},\mathcal{A}_B,p^f,\rho,\hat{r}_B,\gamma_B)$, where $f\in\mathcal{W}_A$ is the leader's policy, $p^f(s'|s,b):=\E_{a\sim f(s)}[\hat{p}(s'|s,a,b)]$, and $\hat{r}_B:\mathcal{S}\x\mathcal{B}\to\R$.
Notably, $R_B^*(f)$ is its optimal policy.
Then, the problem of guiding the follower's aim to the leader's objective is expressed as
\begin{align}
	\max_{f\in\mathcal{W}_A} \E^{fR_B^*(f)}\left[\sum_{t=0}^{\infty}\gamma_A^t\hat{r}_A(s_t,a_t,b_t)\right],\label{eq:PT_obj}
\end{align}
where $\hat{r}_A$ is the leader's reward function that represents the desired follower's behavior intended by the leader.
$f^*$ is the optimal solution of \cref{eq:PT_obj}, and the state-action sequences generated by the corresponding follower's optimal policy $R_B^*(f)$ maximize not only the follower's expected cumulative discounted reward but also that of the leader's.
This can be viewed as the follower attaining the leader's desired behavior when the follower achieves its own learning goal.


The optimal solution of \cref{eq:PT_obj} is given by the PO policy $f$ that maximizes $\mathcal{L}[V_{A}^{f\dag}] = \sum_{s \in \mathcal{S}} \rho(s) V_{A}^{f\dag}(s)$.
Therefore, our proposed approach can be applied to this problem when the transition function is known and the follower's best response can be computed. 
Although the optimality of the obtained solution is not guaranteed unless $\gamma_A = 0$, our analysis guarantees the monotone improvement of the leader's policy, which is of great importance in practice.
In contrast, existing methods trying to obtain the SSE policy may be inadequate because it is usual that the follower is not myopic, and hence, the SSE policy does not necessarily exist.


\begin{thebibliography}{23}
\providecommand{\natexlab}[1]{#1}
\providecommand{\url}[1]{\texttt{#1}}
\expandafter\ifx\csname urlstyle\endcsname\relax
  \providecommand{\doi}[1]{doi: #1}\else
  \providecommand{\doi}{doi: \begingroup \urlstyle{rm}\Url}\fi

\bibitem[Behzadan and Munir(2017)]{Behzadan2017-vs}
Vahid Behzadan and Arslan Munir.
\newblock Vulnerability of deep reinforcement learning to policy induction attacks.
\newblock In \emph{Machine Learning and Data Mining in Pattern Recognition}, pages 262--275. Springer International Publishing, 2017.

\bibitem[Bucarey et~al.(2019)Bucarey, Della~Vecchia, Jean-Marie, and Ord{\'o}{\~n}ez]{Bucarey2019-bf}
V{\'\i}ctor Bucarey, Eugenio Della~Vecchia, Alain Jean-Marie, and Fernando Ord{\'o}{\~n}ez.
\newblock Stationary strong stackelberg equilibrium in discounted stochastic games.
\newblock Technical Report RR-9271, INRIA, May 2019.

\bibitem[Bucarey et~al.(2022)Bucarey, Vecchia, Jean-Marie, and Ordo{\~n}ez]{Bucarey2022-ro}
V{\'\i}ctor Bucarey, Eugenio~Della Vecchia, Alain Jean-Marie, and Fernando Ordo{\~n}ez.
\newblock Stationary strong stackelberg equilibrium in discounted stochastic games.
\newblock \emph{IEEE Trans. Automat. Contr.}, 68\penalty0 (9):\penalty0 5271--5286, November 2022.

\bibitem[Fiez et~al.(2020)Fiez, Chasnov, and Ratliff]{Fiez2020-lu}
Tanner Fiez, Benjamin Chasnov, and Lillian Ratliff.
\newblock Implicit learning dynamics in stackelberg games: Equilibria characterization, convergence analysis, and empirical study.
\newblock In Hal~Daumé Iii and Aarti Singh, editors, \emph{Proceedings of the 37th International Conference on Machine Learning}, pages 3133--3144.  2020.

\bibitem[Harada et~al.(2006)Harada, Sakuma, and Kobayashi]{Harada2006pdm}
Ken Harada, Jun Sakuma, and Shigenobu Kobayashi.
\newblock Local search for multiobjective function optimization: pareto descent method.
\newblock In \emph{Proceedings of the 8th Annual Conference on Genetic and Evolutionary Computation}, GECCO '06, page 659–666, New York, NY, USA, 2006. Association for Computing Machinery.

\bibitem[Hu and Wellman(2003)]{Hu2003-vy}
Junling Hu and Michael~P. Wellman.
\newblock Nash q-learning for general-sum stochastic games.
\newblock \emph{J. Mach. Learn. Res.}, 4\penalty0:\penalty0 1039–1069, dec 2003.
\newblock ISSN 1532-4435.

\bibitem[Huang and Zhu(2019)]{Huang2019-bh}
Yunhan Huang and Quanyan Zhu.
\newblock Deceptive reinforcement learning under adversarial manipulations on cost signals.
\newblock In \emph{Decision and Game Theory for Security}, pages 217--237. Springer International Publishing, 2019.

\bibitem[Jean-Marie et~al.(2022)Jean-Marie, López, Vecchia, and Ordoñez]{Jean-Marie2022-td}
Alain Jean-Marie, Víctor~Bucarey López, Eugenio~Della Vecchia, and Fernando Ordoñez.
\newblock Stationary strong stackelberg equilibrium in discounted stochastic games.
\newblock Dynamic Games and Applications Seminar, October 2022.
\newblock URL \url{https://www-sop.inria.fr/members/Alain.Jean-Marie/Talks/13Oct2022/Presentation_GERAD_2022.pdf}.

\bibitem[Kao et~al.(2022)Kao, Wei, and Subramanian]{Kao2022-va}
Hsu Kao, Chen-Yu Wei, and Vijay Subramanian.
\newblock Decentralized cooperative reinforcement learning with hierarchical information structure.
\newblock In Sanjoy Dasgupta and Nika Haghtalab, editors, \emph{Proceedings of The 33rd International Conference on Algorithmic Learning Theory}, pages 573--605. 2022.

\bibitem[Kononen(2004)]{Kononen2004-xe}
V~Kononen.
\newblock Asymmetric multiagent reinforcement learning.
\newblock In \emph{{IEEE/WIC} International Conference on Intelligent Agent Technology, 2003. {IAT} 2003}, volume~2, pages 105--121. IEEE Comput. Soc, 2004.

\bibitem[Rakhsha et~al.(2020)Rakhsha, Radanovic, Devidze, Zhu, and Singla]{Rakhsha2020-sc}
Amin Rakhsha, Goran Radanovic, Rati Devidze, Xiaojin Zhu, and Adish Singla.
\newblock Policy teaching via environment poisoning: Training-time adversarial attacks against reinforcement learning.
\newblock In \emph{Proceedings of the 37th International Conference on Machine Learning}, pages 7974--7984. 2020.

\bibitem[Shapley(1953)]{Shapley1953-me}
L~S Shapley.
\newblock Stochastic games.
\newblock \emph{Proc. Natl. Acad. Sci. U. S. A.}, 39\penalty0 (10):\penalty0 1095--1100, October 1953.

\bibitem[Solan(2022)]{Solan2022-ce}
Eilon Solan.
\newblock \emph{A Course in Stochastic Game Theory}.
\newblock Cambridge University Press, May 2022.

\bibitem[Sun et~al.(2020)Sun, Huo, and Huang]{Sun2020-bo}
Yanchao Sun, Da~Huo, and Furong Huang.
\newblock {Vulnerability-Aware} poisoning mechanism for online {RL} with unknown dynamics.
\newblock In \emph{9th International Conference on Learning Representations, {ICLR} 2021, Virtual Event, Austria, May 3-7, 2021}, September 2020.

\bibitem[Sutton and Barto(2018)]{Sutton2018-jf}
Richard~S Sutton and Andrew~G Barto.
\newblock \emph{Reinforcement Learning, second edition: An Introduction}.
\newblock MIT Press, November 2018.

\bibitem[Vu et~al.(2022)Vu, Alumbaugh, Ching, Ding, Mahajan, Chasnov, Burden, and Ratliff]{vu2022stackelberg}
Quoc-Liem Vu, Zane Alumbaugh, Ryan Ching, Quanchen Ding, Arnav Mahajan, Benjamin Chasnov, Sam Burden, and Lillian~J Ratliff.
\newblock Stackelberg policy gradient: Evaluating the performance of leaders and followers.
\newblock In \emph{ICLR 2022 Workshop on Gamification and Multiagent Solutions}, 2022.

\bibitem[Zhang et~al.(2020{\natexlab{a}})Zhang, Chen, Huang, Li, Yang, Zhang, and Wang]{Zhang2020-hw}
Haifeng Zhang, Weizhe Chen, Zeren Huang, Minne Li, Yaodong Yang, Weinan Zhang, and Jun Wang.
\newblock Bi-level actor-critic for multi-agent coordination.
\newblock \emph{Proc. Conf. AAAI Artif. Intell.}, 34\penalty0 (05):\penalty0 7325--7332, April 2020{\natexlab{a}}.

\bibitem[Zhang and Parkes(2008)]{Zhang2008-ll}
Haoqi Zhang and David Parkes.
\newblock Value-based policy teaching with active indirect elicitation.
\newblock \url{https://dash.harvard.edu/bitstream/handle/1/4039771/Zhang_Value.pdf?sequence=2&isAllowed=y}, 2008.
\newblock Accessed: 2023-2-6.

\bibitem[Zhang et~al.(2009)Zhang, Parkes, and Chen]{Zhang2009-oz}
Haoqi Zhang, David~C Parkes, and Yiling Chen.
\newblock Policy teaching through reward function learning.
\newblock In \emph{Proceedings of the 10th {ACM} conference on Electronic commerce}, EC '09, pages 295--304, New York, NY, USA, July 2009. Association for Computing Machinery.

\bibitem[Zhang et~al.(2020{\natexlab{b}})Zhang, Ma, Singla, and Zhu]{Zhang2020-ai}
Xuezhou Zhang, Yuzhe Ma, Adish Singla, and Xiaojin Zhu.
\newblock Adaptive {Reward-Poisoning} attacks against reinforcement learning.
\newblock In \emph{Proceedings of the 37th International Conference on Machine Learning}, pages 11225--11234. 2020{\natexlab{b}}.

\bibitem[Zhao et~al.(2023)Zhao, Zhu, Jiao, and Jordan]{Zhao2023-jc}
Geng Zhao, Banghua Zhu, Jiantao Jiao, and Michael Jordan.
\newblock Online learning in stackelberg games with an omniscient follower.
\newblock In \emph{Proceedings of the 40th International Conference on Machine Learning}, pages 42304--42316. 2023.

\bibitem[Zheng et~al.(2022)Zheng, Fiez, Alumbaugh, Chasnov, and Ratliff]{Zheng2022-xj}
Liyuan Zheng, Tanner Fiez, Zane Alumbaugh, Benjamin Chasnov, and Lillian~J Ratliff.
\newblock Stackelberg actor-critic: Game-theoretic reinforcement learning algorithms.
\newblock \emph{Proc. Conf. AAAI Artif. Intell.}, 36\penalty0 (8):\penalty0 9217--9224, June 2022.

\bibitem[Zhong et~al.(2023)Zhong, Yang, Wang, and Jordan]{Zhong2023-fn}
Han Zhong, Zhuoran Yang, Zhaoran Wang, and Michael~I Jordan.
\newblock Can reinforcement learning find {Stackelberg-Nash} equilibria in {General-Sum} markov games with myopically rational followers?
\newblock \emph{J. Mach. Learn. Res.}, 24\penalty0 (35):\penalty0 1--52, 2023.

\end{thebibliography}
\end{document}